\documentclass[preprint,12pt]{aastex}

\newcommand{\msb}{$M_{\odot}$~}

\shorttitle{Nucleosynthesis in N Stars}

\shortauthors{Abia et al. }

\begin{document}

\title{$s$-Process Nucleosynthesis in Carbon Stars}

\author{C. Abia\altaffilmark{1} \& I. Dom\'\i nguez\altaffilmark{1}}
\affil{Dept. F\'\i sica Te\'orica y del Cosmos, Universidad de
Granada, E-18071 Granada, Spain} \email{cabia@ugr.es; inma@ugr.es}

\author{R. Gallino\altaffilmark{2}}\affil{Dipartimento di Fisica
Generale, Universit\'a di Torino and Sezione INFN di Torino, Via
P. Giuria 1, 10125 Torino, Italy} \email{gallino@ph.unito.it}

\author{M. Busso\altaffilmark{3}}\affil{Dipartimento di Fisica,
Universit\`a di Perugia, Via Pascoli, 06123 Perugia, Italy}
\email{Maurizio.Busso@fisica.unipg.it}

\author{S. Masera\altaffilmark{4}}\affil{Dipartimento di Fisica
Generale, Universit\`a di Torino, Via P. Giuria 1, 10125 Torino,
Italy} \email{smasera@to.infn.it}

\author{O. Straniero\altaffilmark{5}}\affil{Osservatorio
Astronomico di Collurania, I-64100 Teramo, Italy}
\email{straniero@astrte.te.astro.it}

\author{P. de Laverny\altaffilmark{6}}\affil{Observatoire de la
Cote d'Azur, Dpt. Fresnel UMR 6528, Nice, France} 
\email{laverny@obs-nice.fr}

\author{B. Plez\altaffilmark{7}}\affil{Universit\'e de
Montpellier 2, Montpellier, France}
\email{plez@graal.univ-montp2.fr}

\and

\author{J. Isern\altaffilmark{8}}\affil{Institut d'Estudis
Espacials de Catalunya (CSIC), Barcelona, Spain}
\email{isern@ieec.fcr.es}

\begin{abstract}

We present the first detailed and homogeneous analysis of the
$s$-element content in Galactic carbon stars of N-type.
Abundances of Sr,Y, Zr (low-mass $s$-elements, or ls) and of Ba,
La, Nd, Sm and Ce (high-mass $s$-elements, hs) are derived using
the spectral synthesis technique from high-resolution spectra.
The N-stars analyzed are of nearly solar metallicity and show
moderate $s$-element enhancements, similar to those found in S
stars, but smaller than those found in the only previous similar study 
(Utsumi 1985), and also smaller than
those found in supergiant post-AGB stars. This is in agreement
with the present understanding of the envelope $s$-element
enrichment in giant stars, which is increasing along the spectral
sequence
M$\rightarrow$MS$\rightarrow$S$\rightarrow$SC$\rightarrow$C during
the AGB phase. We compare the observational data with recent
$s$-process nucleosynthesis models for different metallicities
and stellar masses. Good agreement is obtained between low mass
AGB star models ($M\la 3 M_\odot$) and $s$-elements observations.
In low mass AGB stars, the $^{13}$C($\alpha, n)^{16}$O reaction
is the main source of neutrons for the $s$-process; a moderate
spread, however, must exist in the abundance of $^{13}$C that is burnt in
different stars. By combining information deriving from the
detection of Tc, the infrared colours and the theoretical
relations between stellar mass, metallicity and the final C/O
ratio, we conclude that most (or maybe all) of the N-stars
studied in this work are intrinsic, thermally-pulsing AGB stars;
their abundances are the consequence of the operation of third
dredge-up and are not to be ascribed to mass transfer in binary
systems.

\end{abstract}


\section{Introduction}

It is known that the chemical composition of the interstellar
medium is oxygen-rich. Hence, the overwhelming majority of the
stars are formed with a carbon to oxygen ratio lower than unity,
and most of them do not change this property during their
evolution. However, there are exceptions: several classes of
stars are known, whose carbon to oxygen ratio in the envelope is
larger than unity (by number of atoms). These stars are named
carbon (C) stars. The existence of carbon stars must be related
to some specific mechanism, acting on a limited population of
objects. There are basically three possibilities: i) the carbon
enrichment is the result of a deep mixing process suitable to
pollute the photosphere with the carbon synthesized by shell
He-burning in the stellar interior; ii) it is due to mass transfer
of carbon-rich material after the stellar birth; iii) it dates
back to the star's birth, due to an anomalous composition of the
parental cloud, more enriched in carbon than in oxygen.
Concerning the last hypothesis, so far no such interstellar
clouds have been detected in Galactic disk environments, though
the hypothesis is not completely ruled out for certain low
metallicity stars (Beveridge \& Sneden 1994). In the second case,
the origin of the carbon-rich material is simply moved to another
place, i.e. the carbon-rich primary component of a binary system.
For instance, the stars of the CH spectral type (Luck \& Bond
1991; Vanture 1992a,b,c) most likely owe their carbon enhancement
to the transfer of carbon-rich material from a companion (now a
white dwarf). These stars are usually named {\it extrinsic} 
asymptotic giant branch (AGB) stars to distinguish them from 
those deriving their carbon enhancement from nuclear processing 
in their interiors (hypothesis i), which are called 
{\it intrinsic}. Here, we shall deal with a specific subclass of 
carbon stars, those of spectral type N. We shall see later that most 
probably all of them can be
classified as intrinsic C-rich objects. They are formed through
the mixing into the envelope of newly produced $^{12}$C from He
burning. Stellar evolution then limits the candidates to evolved
stars with masses $1\la M/M_\odot\la 8$, in particular those
ascending AGB. The structure of an
AGB giant is characterized by a degenerate CO core, by two shells
(of H and He) burning alternatively, and by an extended convective
envelope. In the HR diagram, these stars lay close to the
brightest part of the Hayashi line. They become long period
variable of the irregular, semiregular or Mira types, presenting
large mass-loss rates: $10^{-8}$ to $10^{-4}$ M$_\odot/$yr
(Wallerstein \& Knapp 1998). As a consequence, a thick
circumstellar envelope eventually forms, sometimes developing
detached shells. Depending on its chemical composition and
optical thickness, this  circumstellar material can obscure
partially or completely the central star at optical wavelengths
(Knapp \& Morris 1985; Olofsson et al. 1993; Marengo et al. 2001).
Schwarzschild \& H\"arm (1965) early showed  that thermal
instabilities in the He-shell (thermal  pulses, TP) occur
periodically during the advanced phases of AGB evolution.  During
a TP the whole region between the H shell and the He shell
(called ``the He intershell'') becomes convective. After each TP,
the convective envelope penetrates downward dredging-up material
previously exposed to incomplete He-burning conditions. This
phenomenon is called {\it third dredge up} (TDU), and its main
consequence is the increase of the carbon content in the envelope
so that, eventually, the C/O ratio can exceed  unity and the star
becomes a carbon star. In such a way, the carbon content in the
envelope is expected to increase along the spectral sequence
M$\rightarrow$MS$\rightarrow$S$\rightarrow$SC$\rightarrow$C, stars
of spectral class C showing C/O$>1$ (see e.g. Iben \& Renzini
1983; Smith \& Lambert 1990).

Another important consequence of TDU is the  enrichment of the
envelope in {\it s}-elements. The necessary neutrons  for the
$s$-process are released by two reactions:
$^{13}$C$(\alpha,n)^{16}$O, which provides the bulk of the
neutron flux at low neutron densities ($N_n\la 10^7$ cm$^{-3}$),
and $^{22}$Ne$(\alpha,n)^{25}$Mg, which is activated at
temperatures $T\ga 3.0\times  10^8$ K, providing a high peak
neutron density ($N_n\sim 10^{10}$ cm$^{-3}$) and is responsible
of the production of $s$-nuclei controlled by reaction branchings
(see Wallerstein  et  al. 1997; Busso, Gallino, \& Wasserburg
1999 and references therein). The abundances of $s$-nuclei are
known to increase along the above mentioned spectral sequence, as
the star gradually ascends the AGB. Evidence of this was provided
during the last few decades by several studies on AGB stars of
different spectral types:  MS and S stars (C/O$<1$) (Smith \&
Lambert 1985, 1986, 1990); SC stars (C/O$\sim 1$) (LLoyd-Evans
1983; Abia \& Wallerstein 1998, hereafter Paper I) and even
post-AGB supergiants of spectral types A and F  (Van Winckel \&
Reyniers  2000; Reddy, Bakker, \& Hrivnak 1999).

The above studies found consistent enhancements of $s$-elements
with respect to ``normal'' red giants assumed as comparisons (or
with respect to the Sun): from [s/Fe]$\approx +0.3$ in MS giants
to [s/Fe]$>+0.5$ in S stars\footnote{We adopt the usual notation
[X/Y]$\equiv$ log (X/Y)$_{\rm{program star}}-$ log
(X/Y)$_{\rm{comparison star}}$ for the stellar value of any
element ratio X/Y.}. The enrichment seems to continue along the
AGB phase until the planetary nebula ejection. Indeed, post-AGB 
supergiants show high enhancements, [s/Fe]$> +1.0$. In this scenario, 
however, normal
carbon stars (N-type)\footnote{There are carbon stars of R-type
and J-type, which probably owe  their carbon enhancement to a
mechanism different than the TDU. These stars  are not
significantly enhanced in $s$-elements (Dominy 1985; Abia \&
Isern 2000)} have still to find a place. This is so because of
the complex spectra of N stars, which are so crowded with
molecular and atomic absorption features (many of which
unidentified) that abundance analysis has been strongly limited.

The situation has not improved much with the advent of spectrum
synthesis techniques. Indeed, the only abundance studies
available to date are those by Kilston (1975) and Utsumi (1970,
1985), still based on abundance indexes and on low-resolution
photographic spectra, respectively. These works suggested that N
stars were $s$-element rich, showing overabundances by a factor
of ten with respect to the Sun. As far as the AGB phase is
concerned, this figure is not in direct contradiction with the
accepted general scenario of $s$-nuclei enhancement (see e.g.
Busso et al. 1995). More likely, such high production factors
might be difficult to reconcile with the enrichment in post-AGB
supergiants showing similar, or even smaller, abundances.
However, the large observational uncertainties for C-rich red
giants, and the lack of adequate model atmospheres has so far
prevented a solid theoretical interpretation.

In a previous  work (Abia et al.  2001, hereafter Paper  II) we
presented high-resolution spectroscopic  observations for  a
sample of  N stars focusing  our attention on light $s$-elements
(ls), sited at the abundance peak near the neutron-magic number
$N=50$, around the $^{85}$Kr branching point of the $s$-process
path. We showed how the analysis of the abundance ratios of Rb (a
neutron-density sensitive element, see e.g. Beer \& Macklin 1989)
relative to its neighbors (Sr, Y and Zr) yields information on
important details of the $s$-process mechanism operating, and on
the initial stellar mass. We concluded that $s$-processing
suggests low mass stars (LMS, $M\la 3 M_\odot$) as the likely
parents of C(N) giants.

In LMS the major neutron source is $^{13}$C, which burns
radiatively in a tiny layer during the interpulse phase
(Straniero et al. 1995) at relatively low temperatures ($\approx$
8 keV), as a consequence of the formations of a $^{13}$C-rich
{\it pocket} in the intershell region. In the rarer AGB stars of
intermediate mass (IMS) ($M\ga 4 M_\odot$), $^{22}$Ne would
instead be favored as a neutron source, by the higher temperature
in thermal pulses. Because of the very different neutron density
provided by the two neutron-producing reactions, different
compositions are expected from them, especially for the ls
mixture. It is on this basis that in Paper II we drew our
conclusions for the initial masses.

In Paper II we also discussed the $^{12}$C/$^{13}$C ratios
measured in the sample stars, showing that  most of them cannot
be  explained by canonical stellar models on the AGB phase,
requiring probably the operation of an ad-hoc mixing mechanism.
This mechanism is often indicated with the term 'cool bottom
process' (CBP). It is expected to occur in low-mass stars during
the red giant branch and perhaps, also during the AGB phase
(Wasserburg, Boothroyd, \& Sackmann 1995; Nollett, Busso, \&
Wasserburg 2002).

In the present work we have extended our study of $s$-element
nucleosynthesis in N stars to the nuclei belonging to the second
$s$-process peak: Ba, La, Ce, Nd, and Sm. We also re-analyzed the
$s$-element abundances already derived in Paper II, by applying
the spectral synthesis method. Together, these data are compared
with recent models for $s$-processing in AGB stars at the
metallicities relevant  for our  sample stars. Furthermore, using
the infrared properties and Tc content, we discuss the
possibility that our stars be extrinsic carbon stars, concluding
that this is unlikely for several reasons. The structure of the
paper is the following. In $\S~2$ we present the characteristics
of the stars and the spectroscopic observations. Section $\S~3$
describes the method of  analysis and the  sources of error.  Our
abundance results are reported in $\S~4$ together with a
comparison with similar studies in other AGB stars and with
models of AGB nucleosynthesis. Finally, in $\S~5$ we summarize
the main conclusions that can be drawn from this work.

\section{Observations and analysis}

The spectra of our N star sample were collected at the Roque de
los Muchachos (ORM) and Calar Alto (CAHA) observatories in
several observational runs between the years 1998 and 2001. In
the former case we used the 4.2 m William Herschel telescope and
the 2.5 m Nordic telescope with the UES and SOFIN spectrographs,
respectively. The UES gives a maximum resolution of about 50000
between $\lambda=450$ and $\lambda=900$ nm, with some gaps
between orders. SOFIN, working with the third camera
configuration, reaches a maximum resolution of 180000 in 28
orders, covering about the same spectral range but with larger
gaps between orders. At the CAHA observatory we used the 2.2 m
telescope with the FOCES spectrograph which provides a resolution
of about 35000 between $\lambda =360$ and $\lambda= 1000$ nm. In
this case, an almost complete spectral coverage is obtained.
Table 1 shows the sample stars and the epoch of the observations.
Data reduction was performed following the standard procedures
using the IRAF software package. Details on this are quoted in
Abia \& Isern (2000) and in Paper II. In the present work we add
new N stars to the sample studied in Paper II, namely: RV Cyg, S
Aur, SS Vir, T Dra, TT Cyg, U Hya, U Lyr, VY Uma, V429 Cyg and W
CMa.

\subsection{Stellar parameters and luminosities}

The main parameters of the stars were derived in a way similar to
the one of Paper II; they will not be discussed again here (see
that paper for details). For all the stars we adopted a typical
gravity of log $g=0.0$ and a microturbulence of $\xi=2.2$ km
s$^{-1}$. As in previous works, we used the grid of model
atmospheres for carbon stars computed by the Uppsala group (see
Eriksson et al. 1984 for details). The absolute abundances of
CNO, the C/O ratio (which mainly determines the global shape of
the spectrum in carbon stars) and the $^{12}$C/$^{13}$C ratios
were derived as discussed in Paper II. Table 2 shows the final
parameters adopted for each star.

In order to estimate the absolute bolometric magnitudes
M$_{\rm{bol}}$ of our stars (Table 1) we used the empirical
relationships between M$_{\rm{bol}}$  and M$_{\rm K}$ for carbon
stars obtained  by \citet{alk98}. The average K-magnitude values
(the stars studied are variable!) were taken from the {\it Two
Micron Sky Survey} (Neugebauer \& Leigthon 1969), available at
the SIMBAD data base. Actually, though recent catalogs of near
infrared images and point sources have become available (namely:
2MASS, Skrutskie et al. 1997, and DENIS, Epchtein et al. 1999),
our sources are too bright for them, and are saturated. Recent
observations at infrared wavelengths have been collected by the
recent fifth edition of the Catalog of Infrared Observations,
(Gezari et al. 1999); a few of our stars are present in the
compilation, and the colors derived are compatible with those
used here within the uncertainties. K magnitudes were corrected
for interstellar extinction according to the Galactic extinction
model by \citet{are92} with $A_{\rm{K}}/A_{\rm{V}}=0.126$
(Cardelli et al. 1989). Distances were calculated from the {\it
Hipparcos Intermediate Astrometric Data} (van Leeuwen \& Evans
1998). The Hipparcos parallaxes for carbon stars generally show
much lower precision than those calculated for hotter main
sequence stars.
Parallaxes for carbon stars of different spectral types have been
partially reprocessed (Knapp, Pourbaix, \& Jorissen 2001;
Pourbaix, Knapp, \& Jorissen 2002; Bergeat et al. 2001, Bergeat,
Knapik, \& Rutily 2002a, b). However, the bolometric magnitudes
showed in Table 1 have still to be taken with some caution. In
any case, it is remarkable that the overwhelming majority of the
stars have M$_{\rm bol}\la -7.1$, the maximum theoretical
luminosity of an AGB star when its core mass reaches the
Chandrasekhar mass limit, according to core-luminosity relation
by Paczy\'nski (1970). Our sample stars have in average M$_{\rm
bol}\sim -5$, but a considerable number have lower luminosity.
This figure coincides with recent luminosity studies of carbon
stars in the galaxy (Bergeat, Knapik, \& Rutily 2002a) and in the
nearby stellar systems (Westerlund et al. 1991, 1995)

\subsection{The list of lines}

Utsumi (1970) first noted that in N stars the blend effect of
molecular bands is not very serious in the spectral regions
$\lambda \approx 4400-4500$ and $\lambda \approx 4750-4950$
{\AA}. He also found that many lines of $s$-process and
rare-earth elements present in these spectral regions are of
moderate intensity, and therefore they are useful for abundance
analysis. Here, however, we have concentrated our analysis in the
$\lambda=4750-4950$ {\AA} spectral range since the radiation flux
received from many carbon stars below 4500 {\AA} is very limited,
because of the presence of strong C$_3$ molecular absorption
features below that wavelength limit. In Papers I and II we used
the equivalent width versus curve of growth technique to derive
heavy element abundances trough a careful identification and
selection of the lines (see these papers for details). In the
present study however, we decided to use the spectral synthesis
technique because most of the lines from $s$-elements with high
atomic mass (Ba, La, Ce, Nd and Sm, i.e. hs elements) present in
this spectral range are heavily blended. This also allows us to
verify the results of Paper II for the ls elements (Sr, Y, Zr),
re-deriving their abundances independently. This check confirmed
the abundances deduced in Paper II (within the error bars, with
the exception of a few cases, for reasons that are discussed
below). This gives an important confirmation of the previous
analysis technique.

The spectral  window studied, while being less affected by
molecular blends than others, is nevertheless not free from such
problems. Lines from the C$_2$, CH and CN molecules have
therefore to be considered. Transitions from the $^{12}$C$_2$,
$^{12}$C$^{13}$C and $^{13}$C$_2$ (A-X), (B-A), (D-A) and (E-A)
systems were found in the Kurucz database
(cfaku5.harvard.edu/linelists.html). From the work of
\citet{deL98} and Kurucz data, no C$_2$ lines from the Phillips
red system were predicted to be of importance in the studied
spectral ranges. $^{12}$C$^{14}$N,  $^{13}$C$^{14}$N,
$^{12}$C$^{15}$N and $^{13}$C$^{15}$N lines were included. The CN
line lists were prepared along the same lines as the TiO line
list of \citet{ple98}, using data from \citet{ito88},
\citet{kot80}, \citet{cer78}, \citet{reh92}, \citet{pra92},
\citet{pra92b}, \citet{bau88} and \citet{lar83}. Programs by
Kotlar were used to compute wavenumbers of transitions in  the
red bands studied by \citet{kot80}. Lines from the violet system
were found to be very weak but were nevertheless considered. The
absorption by the C$_2$, $^{13}$C$^{15}$N, $^{12}$C$^{15}$N and
$^{13}$C$^{14}$N lines are found to slightly decrease the
theoretical continuum by a few percent without affecting the line
depths. In that spectral range, the main contributor from these
two molecules is $^{12}$C$^{14}$N. We adopted a solar
$^{15}$N/$^{14}$N in all the spectral synthesis. Several CH lines
also play a role in the $4750-4950$ {\AA} spectral range. The
LIFBASE program of \citet{lif99} was used to compute these line
positions and $gf$-values. Excitation energies and isotopic
shifts (LIFBASE provides only line positions for $^{12}$CH) were
taken from the line list of \citet{jor96}. In this way, a very
good fit of CH lines in the solar spectrum could be obtained,
with the exception of very few lines which we removed from the
list. Obviously, the line list also comprises atomic line data
found in the VALD database \citep{kup98} and in Kurucz's data
(CD-ROM No. 23) although, for the more intense  atomic lines
(mostly metals), we derived solar $gf$-values using the Holweger
\& M\"uller (1974) model atmosphere and solar abundances by
Anders \& Grevesse (1989). The $gf$-values derived in such a way
showed an excellent agreement with the solar values derived by
Th\'evenin (1989, 1990). For a sample of about 100 lines in common
we found a mean difference of $0.15\pm 0.2$ dex. Our final line
list includes more than 20000 atomic and molecular lines. Despite
of this, only a few lines were selected as indicators of the hs
element abundances in our stars. These lines are shown in Table
3. The metallicity [M/H] of our sample stars was derived from
selected metallic lines, ``M'' indicating an average value of Ca,
Ti, V, Fe and Co abundances (see Table 3 in Paper II). Using this
average changes only slightly the values [ls/Fe] and [hs/Fe]
derived.

\subsection{Data analysis, peculiar cases, comparison with previous work}

The hs element abundances were derived from spectrum synthesis
fits to the observed spectra. The theoretical spectra were
computed in LTE and the abundance analysis was done line-by-line
relative to the star WZ Cas (C9,2J;SC7/10) which was used, as in
Paper II, as the reference star. WZ Cas is a carbon star with a
C/O ratio very close to unity with near solar metallicity and
with very small or no $s$-element enhancement: [h/Fe]$=0.02\pm
0.20$, where h (heavies) includes Sr, Y, Zr, Nb, Ba, La, Ce and
Pr (see Lambert et al. 1986; Abia \& Isern 2000). This
comparative analysis might reduce systematic errors due to wrong
$gf$-values and possible departures from LTE \citep{tom83}. All
the abundance ratios derived (see below) are relative to this
star. The final abundance derived for each element is the mean of
the abundances derived from each line, except for Ba whose
abundance is derived from only one line. In deriving these means,
we consider twice the abundances derived from particular lines
which we believe are better abundance indicators for a given
element because of their larger sensitivity to abundance
variations and/or because they are placed in a less crowded
region. These lines are: the $\lambda 4809$ LaII, $\lambda 4773$
CeII and $\lambda 4811$ NdII lines (see Table 3).

Figure 1 shows an example of a theoretical fit to the observed
spectrum of the star Z Psc in the region between $4800-4830$
{\AA}, where most of the available $s$-element lines are located.
Figure 2 is a similar example in the region around the
4934 {\AA} BaII line. It can be seen in both figures that the
observed spectra are generally well fitted by spectral synthesis,
despite the many features still not accounted for. This last fact
is a clear evidence of the remaining incompleteness of the line
list used, despite the several thousands of lines included. 

As noted above, in the present work we have re-derived with the
spectral synthesis technique the abundances of Sr, Y and Zr
obtained in Paper II; this was permitted by the fact that most of
the available lines for these elements are located in the same
spectral range 4750-4950 {\AA} studied here (see Table
3 in Paper II). We find a good agreement (within typical error
bars) with the values found for these elements in Paper II;
hence, our conclusions are not modified by the present analysis.
Only for one star, V460 Cyg, the abundances of Sr, Y and Zr are
significantly modified. The reason of this lays in a bad
interpolation done in Paper II over the grid of model atmosphere
parameters for this star. There is also a modification of the Rb
and Zr abundances in Z Psc with respect to Paper II. The Rb
abundance shown there (Table 2) is just a misprint, the correct
value being [Rb/M]$=0.6$ instead of 0.9. The revised [Zr/M] ratio
here is $+$1.0 dex, which is, nevertheless, in agreement within
error bars with the previous value in Paper II ($+$0.8 dex). In
the present work we have also been able to estimate the Sr
abundance in this star using the Sr I line at 4811.88
{\AA}: we recall that the Sr abundance in Z Psc was not derived
in Paper II.

In Paper II we also discussed the increased difficulties found in
the analysis of certain N stars, in particular IY Hya, VX Gem and
V CrB (see that paper for details). In fact, the abundance
results in such stars were considered with a lower weight in
comparing with theoretical models. Unfortunately, here the list of
{\it difficult} stars has to be extended. Indeed, T Dra, U Lyr
(both Mira variables), S Aur and V429 Cyg show spectra so smeared
out that we had no choice but decide to give up their analysis.
For SS Vir and V Oph (also Mira variable), the spectra are also
smeared out, and in fact they provide an insufficient agreement
between observed and theoretical spectra. In particular, all
Miras in our sample show difficult spectra. It is not clear why
spectral lines in the Mira variables look in this way. Mira
variables have generally thick circumstellar dust shells, whose
contamination due to thermal emission by dust might be
responsible for this problem. The gas motions due to pulsation
might contribute to the broadening of spectral lines. Finally,
the star V781 Sgr shows a weak central emission at the core of
the Na D lines. This is usually interpreted as due to the
presence of a chromosphere. The influence of a chromosphere in
the derivation of abundances in carbon stars is uncertain and is
beyond the scope of this work; therefore, we have included also
this star in the group of {\it difficult} N stars. Summing up,
the abundances derived for IY Hya, VX Gem, V CrB, SS Vir, V Oph
and V781 Sgr have to be considered with considerable caution. To
emphasize this, in the figures below, we have plotted with
smaller symbols the abundance ratios derived for them.

The stellar atmosphere parameters ($T_{\rm{eff}}$, gravity, etc)
for N stars are difficult to determine with accuracy (Bergeat,
Knapik, \& Rutily 2001, 2002a). Note for instance that they are
variable stars, therefore the effective temperature changes during the
pulsation. The microturbulence in variable stars depends on the
wavelength in the spectrum  (due to differences in depth of
formation), and changes with the pulsational phase. Furthermore,
the existence of strong waves, shock fronts, and inhomogeneities
is the rule rather than the exception in their atmosphere (see
Gustafsson \& J\o rgensen 1994 for discussion). In Table 4 we
show the sensitivity of the abundances derived to typical
uncertainties in the atmosphere parameters. Since most of the
lines used in the present work correspond to singly ionized
elements (see Table 3), an important source of uncertainty is the
error in the gravity. The abundances derived are rather sensitive
to the microturbulence parameter, indicating some saturation of
the lines used in the analysis. However, the hs element
abundances are mostly insensitive to uncertainties in
metallicity, CNO absolute abundances and the $^{12}$C/$^{13}$C
ratio. Certainly, deviations from LTE might play a significant
role, although they are reduced in a differential analysis as is
done here. If we consider the uncertainty due to the
placement of the spectral continuum ($\sim 5\%$), adding
quadratically all the sources (non systematic) of error, we found a typical
uncertainty in the absolute abundances of about $\pm 0.2$ dex for
Ba, $\pm 0.4$ dex for La, Nd, and Sm and $\pm 0.45$ dex for Ce.

Furthermore, we have to consider the random error in the
abundances of elements represented only by few lines, as in our
case. We can estimate this random error only in stars where we
derived individual element abundances from more than three lines.
In this case we found a dispersion in the range 0.1 $-$ 0.3 dex.
Nevertheless, when computing the [hs/M] ratios, errors are lower
because some sources of error affect in the same way the
metallicity (see Table 4 in Paper II).
Adding up all these contributions, we estimate a typical total
error (non systematic) in the [hs/M] ratios ranging from $\pm
0.3$ (Ba) to $\pm 0.45$ dex (Ce). A similar discussion can be
done concerning the uncertainties in the derivation of the ls
element abundances and metallicity (see Paper II for details).
Considering this, we estimate a typical error of about $\pm 0.3$
dex in the [hs/ls] ratios. In Table 5 we show the abundance
ratios with respect to the metallicity derived in our stars,
referred to the comparison star WZ Cas. The abundances in Table 5
represent the first detailed analysis of the $s$-element
composition in N stars. In this table we show the previous
results (revised) obtained in Paper II for the ls elements,
together with the new values derived here for the hs elements,
including some additional stars (see $\S 2$). We confirm our
previous finding that N stars are of near solar metallicity:
$<$[M/H]$>=-0.02\pm 0.10$. Only two stars seem to be of low metal
content (IY Hya and V CrB); however, we remind the difficulties
found in the analysis of these objects (see above). Considering
the abundances of Y and Zr as the only good tracers of the ls
element enhancement\footnote{Because of the difficulties in
deriving Sr abundances, we discard this element when computing
the mean low-mass $s$-element abundances ls (see discussion in
Paper II).}, our objects have $<$[ls/M]$> = +0.67\pm 0.24$. In
the same way, considering Ba, La and Nd as representative of the
hs element enhancement (Ce and Sm are derived only in a few
stars; in many of them we set just upper limits), the mean is
$<$[hs/M] $> =+$0.52$\pm 0.29$. This figure represents moderate,
but significant $s$-element enhancement, which is 
clearly smaller than that obtained previously by \citet{uts85}.
We refer the reader to Paper II for the detailed discussion made
there about the explanation of this difference.

\section{Tc and possible binarity}

The radioactive nucleus $^{99}$Tc ($\tau_{1/2} \sim 2 \times 10^5$
yr) is produced along the $s$-process path. Because the AGB phase
spans only a few half-lives of this element, once dredged-up into
the envelope, it should remain present in a certain percentage
throughout the AGB lifetime.

Therefore, the detection of Tc in the atmosphere of an AGB star
(Merill 1952) is an undeniable evidence of the $s$-process working
{\it in situ} ({\it intrinsic} AGB). When, on the contrary, Tc
presence can be excluded, the star most probably owes its carbon
enrichment to a mass transfer episode from an AGB companion, {\it
extrinsic} AGB. Unfortunately, the detection of Tc in AGB stars
is a difficult task. The most interesting Tc lines are located
around $\lambda\approx 4260$ {\AA}, where AGB stars are not very
bright.

Indeed, in most N stars this spectral region is not accessible to
observations because of the strong flux depression at these
wavelengths. There are, however, a couple of inter-combination Tc
lines that may be used for the analysis. One of them is the TcI
line at $\lambda~5924.47$ {\AA}, which is however weak and not
free from blends. We have extended the analysis of Tc in N stars
made in Paper II using this line in the additional stars included
in this work. We have followed the same technique as described in
Abia \& Wallerstein (1998). In Table 5 we show the results of
this search. As in Paper II, a ``yes'' entry in this table means
that the best fit to the 5924 {\AA} blend is obtained with a
non-zero Tc abundance. A ``no'' entry means that a synthetic
spectrum with no Tc does not significantly differ from another
one with a small Tc abundance and, finally, a ``doubtful'' entry
means that the 5924 {\AA} blend can be fitted better with a small
abundance of Tc, but the Tc absorption in the blend is not so
clearly appreciated as in the case of the Tc ``yes'' stars. If we
consider the ``doubtful'' stars as yielding a detection, which is
probably the case,  Table 5 shows that about $62\%$ of our
targets are Tc-yes N stars and thus, can be considered as {\it
bona fide} intrinsic carbon stars. However, because of the
difficulty in the analysis of this Tc line (weak, and in a very
crowded zone), a non-detection does not necessarily mean that Tc
is absent: we do not have, unfortunately, a clear-cut possibility
of excluding it. Probably, a study of the resonance lines at
$\lambda\sim 4260$ {\AA}, when feasible, could yield more secure
conclusions. Thus, additional tests are necessary to establish
the extrinsic or intrinsic nature for the rest of the sample.

A further help can luckily come from an analysis of infrared
colors. This idea (Jura 1986; Little-Marenin \& Little 1988)
arises from the fact that intrinsic AGB stars (in our case N
stars) are high mass losers and show excess infrared radiation
from circumstellar dust, whereas extrinsic N stars would be
first-ascent red giants with mass loss rates smaller by two
orders of magnitudes at least. This was already used by Jorissen
et al. (1993) to distinguish between extrinsic and intrinsic S
stars. These authors showed that the overwhelming majority of
Tc-deficient S stars have a ratio $R=F(12 \mu m)/F(2.2 \mu
m)<0.1$, where $F(12 \mu$m) and $F(2.2 \mu$m) are the 12 $\mu$m
and 2.2 $\mu$m fluxes from the IRAS Point Source Catalog and from
the Two-Micron Sky Survey, respectively. On the contrary, most
Tc-rich S stars (intrinsic) have $R>0.1$.  We have performed a
similar study in our sample of carbon stars and the result can be
seen in the histogram of Figure 3. We have considered in this
plot the doubtful stars as showing Tc in their spectra, because
of the reasons commented above. From this figure it is obvious
that most of our stars have infrared excess $R>0.1$ and only a
few show lower values (in fact in the very narrow range $\sim
0.08-0.1$). It is clear that, contrary to what happens for
the S stars (see Jorissen et al. 1993), there is no distinction
in the distribution between Tc-yes and Tc-no carbon stars. We
might then conclude that probably all carbon stars in our sample
are in fact of intrinsic nature and the quoted Tc-no carbon stars
are just a consequence of the difficult and uncertain analysis of
the Tc I $5924$ {\AA} blend on which this study is
based\footnote{Note that, as mentioned in Paper II, in some stars
we were able to study simultaneously the stronger resonance TcI
lines at $\lambda\sim 4260$ {\AA}. In all the stars where this was
possible, we confirmed the detection found from the 5924 {\AA}
blend.}.

Similarly to what happens in S stars, a difference between Tc-yes
and Tc-no carbon stars might be also apparent in their
temperature class. The large uncertainty in the derivation of the
effective temperature in carbon stars does not allow us to address
this question directly from the estimated $T_{\rm{eff}}$ values.
The point is that Tc-yes carbon stars should be cooler and more
evolved and hence should (on average) have more circumstellar
matter to account for the infrared excess. This question can be
addressed by studying the K$-[12]$ vs. K$-[25]$ color-color
diagram. The color index K$-[i]$ is defined as
K$-2.5$~log$(620/F(i))$, $F(i)$ being the (non color corrected)
flux in the $i$ band taken from the IRAS Point Catalog Source.
Again, we do not find a distinction between Tc-yes and Tc-no stars
(see Figure 4). All the stars are located in the region of black
bodies with $T\la 3000$ K. These temperatures are indeed expected
in evolved stars during the AGB phase of evolution (note than
some stars have black body temperatures lower than 2000 K).
Higher black body temperatures would instead be expected in
extrinsic (RGB) carbon stars.

A better discrimination will be possible in this field, when
extensive colors for C-rich and O-rich AGB stars from modern
infrared cameras will become available, providing data with a
higher precision than possible from IRAS (see preliminary results
by Marengo et al. 1999; Busso et al. 2001b).

Finally, we can also address the problem of the nature of our
sample stars (i.e. those being extrinsic or intrinsic AGBs)
through a comparison with theoretical models of element enrichment
in the two classes of objects, to look for parameters helping in
the discrimination. There are several questions for which we can,
in this way, find useful answers. Is it possible, at nearly solar
metallicity, to produce an extrinsic carbon star (C/O$>1$) from
the transfer of carbon-rich material in a binary system (most
probably through the wind accretion mechanism, Jorissen \& Mayor
1988)? What should be the C/O ratio in the transferred material so
that the mass-receiving companion remains C-rich? How much these
requirements depend on the stellar mass and on the metallicity of
the accreting star and of the donor? We shall come again on these
problems in the next section, where we present a comparison of
our observations with the results of nucleosynthesis and mixing
models for intrinsic and extrinsic AGB stars.

\section{Comparison with models of AGB nucleosynthesis}

In order to understand the way in which the observed abundances
were created and subsequently mixed into the stellar convective
envelopes, we made use of the same stellar and nucleosynthesis
models already described in Paper II. They start from full stellar
evolution calculations, computed with the FRANEC evolutionary
code, spanning the metallicity range from solar to 1/20 solar,
and the mass range from 1.5 to 7 $M_{\odot}$ (Straniero et al.
1997, 2000).

For our purposes, we adopt the 1.5 $M_{\odot}$ model as
representative of LMS, since in Paper II we already concluded that
most N stars are of low mass ($M\la 3 M_{\odot}$). Notice that the
theoretical expectations for a 3  $M_{\odot}$ stellar model mimic
quite closely the  1.5 $M_{\odot}$ model: differences in
$s$-element logarithmic abundances are found to be below 0.1 dex.
We have computed models without mass loss, from which an
asymptotic value of the parameter $\lambda =\Delta M({\rm
TDU})/\Delta M_{\rm H}$ has been derived, namely 0.24 for the 1.5
$M_{\odot}$ model.
On this basis a post-processing has been performed to derive the
detailed nucleosynthesis products in the He intershell and the
chemical modifications of the envelope (Gallino et al. 1998). The
duration of the thermally pulsing AGB phase is mainly determined
by the assumed mass loss rate. Hence, mass loss was included in
the nucleosynthesis calculations trough the Reimers's (1975)
phenomenological law, choosing the value 0.7 for the free
parameter $\eta$. In each interpulse/pulse cycle we compute the
detailed nucleosynthesis pattern that is established in the
intershell zone, induced first by the neutron release from
radiative $^{13}$C burning in the interpulse phase, and then by
the activation of the $^{22}$Ne neutron source in the convective
instability of the He shell. One cycle after the other, we follow
the penetration of envelope convection through TDU that mixes
part of the processed material to the envelope; the whole
sequence is repeated till our models experience the end of TDU.
The procedure assumes that some form of proton penetration below
the convective border can occur at dredge up, polluting the
radiative He- and C-rich layers, so that at the restart of
H-shell burning the main neutron source $^{13}$C can be formed.
For the problems and the uncertainties involved in this
assumption, as well as for the techniques used in the
computations of neutron captures and the nuclear parameters used,
see Paper II and Gallino et al. (1998). There, in particular, we
described the way in which our model sequences are created, and
the efficiency of $^{13}$C burning assumed. As in Paper II, we
shall refer with the term ST ({\it standard}) to models in which
the mass of $^{13}$C burned per cycle is 4$\times$10$^{-6}$
$M_{\odot}$ in LMS. Then this reference abundance is scaled
upward and downward to span a wide range of possibilities in the
nucleosynthesis efficiency.

Figures 5 to 7 show a synthesis of our predictions: they present,
as a function of the initial metallicity [M/H], the average
photospheric abundance ratios [hs/M], [ls/M], [hs/ls] at C/O =1,
in our 1.5 $M_{\odot}$ model, assumed as representative of the
LMS class.

Inspection of Figures 5, 6 and 7 provides a quite satisfactory
agreement between observations and theoretical models, inside the
wide spread of neutron release efficiency considered and taking
into account the observational uncertainties of N stars. This is
in agreement with previous findings on the average $s$-process
efficiency by Busso et al. (2001a), derived by other classes of
AGB stars and their descendants at various metallicities.

A rather comprehensive picture of the $s$-enrichment in N stars
is provided by plots in which observations and theory are
compared using the $s$-element enhancement (indicated by e.g. ls)
and a parameter sensitive to the neutron exposure (e.g [hs/ls], as
shown in Busso et al. 1995, 2001a). This kind of comparison is
performed in Figure 8, where N stars are shown together with
other known classes of intrinsic and extrinsic AGB stars. As the
figure shows, the spread in the $s$-process efficiency (as
monitored by the N($^{13}$C)/N($^{56}$Fe) ratio in the pocket),
is in this way constrained better, being rather limited. Another
result is that N stars are very close to S stars for what
concerns the efficiency of mixing into the envelope. In the
models, it is indeed often sufficient to make one extra pulse to
change the composition of an S star into that of a N giant,
establishing a C/O ratio larger than unity. This has large
consequences on the appearance of the atmosphere and of the
circumstellar envelope (chemistry, colors, type of dust formed
all change considerably), but the corresponding change in the
$s$-process enhancement may be rather limited. Since post-AGB stars
usually show larger $s$-element enhancements than those derived
here in N-stars (see $\S~1$), and because our sample stars show
a C/O ratio very close to one, we might speculate the stars analyzed
here have just become carbon stars. As the star evolves in the AGB 
phase and TPs and TDUs continue operating, more carbon and $s$-elements are 
dredged-up into the envelope. Eventually the C/O ratio exceeds unity by a 
such large amount that the star becomes obscure at optical wavelengths  
due to the formation of dust. This scenario might explain the apparent 
{\it gap} in the level of $s$-element enhancement and C/O ratio observed 
between C(N) stars and post-AGB stars: we cannot ``see'' carbon stars 
with a C/O ratio largely exceeding unity. We have to outline
once again that our conclusions are for the moment somewhat
hampered by the still large observational uncertainties for N
stars, whereas typical error bars for the other classes of
$s$-enriched stars in the Galactic disc are of in general lower
(0.20 to 0.25 dex, see e.g. Busso et al. 2001a).

The interpretation of the N star abundances presented so far in
the framework of LMS evolution along the AGB is confirmed, and
even strengthened, by an analysis of the detailed element
distributions in the observed stars, from the data of Table 5. In
order to show this we have plotted in Figure 9 a sample of fits
to individual stars, obtained by comparing their observed
composition with that yielded by the models in the photosphere,
at C/O =1. The plots show a complete agreement, and we believe
this provides a full confirmation that most, if not all, N stars
of the Galactic disk are indeed of low initial mass. In fact,
from the discussion presented in Paper II we know a priori that
plots made similarly to Figure 9, but using models from IMS,
would be often incapable of reproducing at least the ``ls'' zone,
yielding in general a too high Rb/Sr ratio. This conclusion for
the mass of carbon stars  is in line with the present independent
understanding of the different classes of carbon stars, provided
by astrometric, kinematic, and photometric criteria (Bergeat,
Knapik, \& Rutily 2002a, b).

\subsection{Can any of our sources be an extrinsic C-star?}

Using the models outlined in the previous subsection, we can add
something to the discussion of the nature of our sources. This
can be done on an independent basis with respect to the (already
analyzed) presence of Tc and infrared properties. Having in mind
the model compositions discussed above, we can ask ourselves
which predictions for $s$-element and carbon abundances would be
derived in case some of our sample stars were extrinsic AGBs. In
order to study this possibility, we have first to consider that
the resulting C/O ratio after mass transfer is constrained by the
observed C/O value, which is very close to 1. Then we must take
into account the original C and O content in the companion. This
depends on metallicity, and at the moment of mass transfer it has
probably not changed from the original CNO composition. Indeed,
the long duration of core H burning (that precedes any
significant abundance change at the surface) makes this phase the
most likely for the occurrence of mass accretion. Subsequently,
partial $^{12}$C dilution is produced by the first dredge up
(FDU), during the RGB phase; this also increases the abundance of
$^{13}$C ($^{16}$O is only slightly reduced by FDU).

For a star reaching the RGB with a solar mixture of CNO isotopes,
FDU would change CN abundances by a significant fraction (e.g.
$^{12}$C would decrease by 13\% to 40\% in solar metallicity
models spanning the range 1-5 $M_\odot$). Its impact is however
marginal when the star has a high carbon enhancement at the
surface. Furthermore, in case of a lower-than-solar metallicity,
one must take into account that the initial carbon abundance
decreases with Fe, whereas [O/Fe] shows a concomitant
enhancement. One can adopt general rules for this enhancement,
deducing it from observations of large databases of dwarf stars
at different metallicities, e.g. as described in Israelian et al.
(2001). Note that while [O/Fe] vs. [Fe/H] still remains
controversial at [Fe/H]$<-1$, this does not affect our
discussion, because the typical metallicity of our sample stars
is much higher.

We then modeled mass transfer by diluting the AGB envelope
composition, allowing for different ``mixing histories'', which
take into account the different possibilities described above. In
all our attempts we found that, for metallicities not far from
solar, it is impossible to produce an extrinsic carbon star. This
gives a strong support to the indications already obtained from
the observational criteria.

To make this result easily understood, we can express the mass
transferred by means of a semi-analytic formula. If mass transfer
occurs while the secondary component is on the main sequence, the
material accumulates onto a thin radiative region, whose chemical
composition becomes that of the accreted material from the AGB
companion and is therefore characterized by a C/O ratio in excess
of unity. At FDU, a homogeneous composition is established by
mixing the accreted material over more internal layers, where the
original chemical composition, partly modified by H burning, has
been preserved. Then, we can define a dilution factor $f$ as the
ratio of the transferred mass ($ M^{\rm{tr}}_{\rm AGB}$) to the
fraction of the envelope having the original composition ($
M^{\rm env,ini}_{\rm comp}$). By imposing that, in the photosphere
of the secondary star, the C/O ratio remains equal to 1 even after
FDU, one can see that:

$$ f =(1 - 0.28/10^{-0.4[\rm{Fe/H}]}) ((\rm{C/O})^{lastTDU}_{AGB}-1)^{-1}$$

Here we have assumed that mass transfer occurs at the last TDU
episode, in order to allow the maximum possible carbon enhancement
of the secondary component, thus favoring the formation of an
extrinsic carbon star. According to the FRANEC models, TDU ceases
when the envelope mass drops below a critical limit, namely
$\sim$0.5 $M_\odot$; from then on, while mass loss can further
decrease the envelope mass, its C/O ratio remains frozen. In Table
6 we report, for the 1.5 $M_\odot$ model at different
metallicities (column 1), the final C/O ratio of the intrinsic
AGB star (primary component) at the last thermal pulse that
experienced TDU ((C/O)$^{\rm{last TDU}}_{\rm{AGB}}$) and the corresponding
dilution ratio $f$, as derived by means of the formula reported
above. Obviously, there is a maximum limit in the transferred
mass, which corresponds to the accretion of the whole residual
mass of the AGB envelope (i.e. $\sim$0.5 $M_\odot$). In order to
give an estimate for such a limit, we further assumed that the
mass of the companion star, after the mass accretion, is the
lowest compatible with the age of the Galactic disk, let us say
$\sim$1 $M_\odot$. Again, this is intended to favor carbon
enrichment, as it roughly corresponds to the minimum mass of the
convective envelope at the FDU. Correspondingly, in the most
favorable conditions, we obtain an upper limit for the dilution
factor $f$ of 1.5. We emphasize that this limit corresponds to a
very extreme case and we expect that a substantially lower value
of the $f$ parameter should be more realistic. For example, if
the mass of the secondary component after the mass accretion
episode were 1.5 $M_\odot$ (corresponding to an initial mass of
about 1 $M_\odot$), the limiting value of the $f$ parameter would
become 0.7.  In any case, from a comparison with the calculated
$f$ values reported in Table 6, it is clear that it would be
practically impossible to form extrinsic carbon stars for solar
metallicities. We suggest that the maximum metallicity of an
extrinsic carbon star should be of the order of [Fe/H] $\sim -$0.4
to $-$0.3.

Thus, even considering the uncertainty in the metallicity of the
stars considered in the present work ($\pm 0.3$ dex), this simple
result prevents the vast majority of them from being considered
as extrinsic carbon stars.

Note that if a carbon star is formed as a consequence of the mass
transfer process, as is indeed the case at lower metallicities,
then the behavior of [ls/Fe] versus [Fe/H] would not be
distinguishable from the trend seen for intrinsic AGBs. This is
related to the fact that mass transfer and the following FDU
behave much like a form of dilution, as in the case of TDU, and
we have no way to distinguish which of the two possible forms of
dilution the $s$-enriched material has undergone (see Busso et
al. 2001a, and in particular Figures 11 and 12 there, for a
discussion of this in the case of Ba and CH stars).

In conclusion, both the observational and theoretical analysis
lead us to conclude that the objects studied here are most
probably intrinsic carbon stars, presently evolving along the
TP-AGB. Exceptions may exists, but this should be tested with
more direct methods, like radial velocity variations and their
effects on the spectra. At the moment, for none of the N stars
studied here there is any evidence of such variations.

\section{Conclusions}

In this paper we have presented for the first time a rather
complete sample of $s$-process abundance measurements in N-type
carbon stars, extended to light and heavy species, across the
$s$-process peaks with neutron numbers $N=50$ and $N=82$. Making
use of a large line list, and of up-to-date model atmospheres for
carbon rich giants, we have deduced the $s$-element abundances
through spectrum synthesis technique. In doing so, we have also
verified previous results (Paper II) for the light $s$-elements
across the $^{85}$Kr branching point of the $s$-path. N stars
turn out to be characterized by s-element abundances very close to, or
slightly higher than, those found for S stars. Compared with the
only analysis performed before (Utsumi 1985), our abundances are
clearly smaller; they are smaller also with respect to those of
post-AGB supergiants (Reddy, Bakker, \& Hrivnak 1999;
Van Winckel \& Reyniers 2000), and this
gives rise to a general picture in which the surface enrichment
continuously increases along the evolutionary sequence producing
MS, S, and N stars, and subsequently yellow post-AGB supergiants
(Reddy et al. 2002). We have compared our data with model envelope compositions
obtained from previously published calculations of AGB
nucleosynthesis and mixing. Good agreement is obtained between
low mass star models and $s$-element observations. Several pieces
of evidence (from the detection of Tc lines and infrared colors,
to the theoretical relations between the initial metallicity, the
mass, and the final C/O ratio) lead us to conclude that most (or
perhaps all) our sample stars are intrinsic TP-AGB stars, so that
their abundances are locally produced by the occurrence of third
dredge-up during the TP-AGB phases, and not generated by mass
transfer in binary systems.

Acknowledgements. Data from the VALD  database at Vienna were
used for the preparation of this paper. K. Eriksson, and the
stellar atmosphere group of the Uppsala Observatory are thanked
for providing the grid of atmospheres. The 4.2m WHT and the  2.5m
NOT are operated on the island of La  Palma by the  RGO in the
Spanish Observatory of the Roque de los Muchachos of the
Instituto de Astrof\'\i sica  de Canarias. This work was  also
based  in part on  observations  collected with  the 2.2m
telescope at the German-Spanish  Astronomical Centre, Calar Alto.
It was partially  supported by the spanish  grants AYA2000-1574,
FQM-292, by  the  Italian   MURST-Cofin2000 project  `Stellar
Observables  of Cosmological  Relevance'  and by  the
French-Spanish  International Program for Scientific
Collaboration, PICASSO HF2000-0087.

\clearpage

\begin{deluxetable}{lcccccc}
\def\n{...}

\tablecolumns{7}
\tablewidth{0pt}
\tablenum{1}
\tablecaption{THE PROGRAM STARS}
\tablehead{\colhead{Star} &
\colhead{Spectral\tablenotemark{1}} &
\colhead{Var.\tablenotemark{1}} &
\colhead{Period\tablenotemark{1}} &
\colhead{M$_{\rm{bol}}$\tablenotemark{2}}& \colhead{Epoch} & 
\colhead{Observatory\tablenotemark{3}}  \\
\colhead{}                             &
\colhead{type}                        &
\colhead{type}                       & \colhead
{(days)}                      & \colhead{} &
\colhead{(JD+2450000)}           & \colhead{}}
\startdata
AQ And     & C5,4        & SR    & 346 & -5.2&  791  & CAHA \\
AW Cyg     & C3,6        & SRb   & 220 & -5.7&  947  & ORM  \\
           &             &       &     &     & 1745  & CAHA \\
EL Aur     & C5,4        & Lb    & \n  & -2.5&  791  & ORM \\
HK Lyr     & C6,4        & Lb    & \n  & -6.0&  947  & ORM \\
           &             &       &     &     & 1015  & CAHA\\
IRC -10397 & N           & \n    & \n  &  \n & 1015  & CAHA \\
           &             &       &     &     & 1745  & CAHA \\
IY Hya     & N           & Lb    & \n  & -5.2&  947  & ORM \\
LQ Cyg     & C4,5        & Lb    & \n  & \n  & 1745  & CAHA \\
RV Cyg     & C6,4        & SRb   & 263 & -7.1& 1745  & CAHA \\
RX Sct     & C5,2        & Lb    & \n  & \n  & 1015  & CAHA \\
S Aur      & C5,4        & SR    & 590 & \n  &  745  & CAHA \\
S Sct      & C6,4        & SRb   & 148 & -4.6& 1015  & CAHA \\
SS Vir     & C6,3        & SRa   & 364 & \n  & 1745  & CAHA \\
SY Per     & C6,4        & SRa   & 476 & -3.1&  791  & CAHA \\
SZ Sgr     & C7,3        & SRb   & 100 & -2.1& 1015  & CAHA \\
T  Dra     & C6,2        & Mira  & 421 & -2.0&  791  & CAHA \\
TT Cyg     & C5,4        & SRb   & 118 & -4.6& 1015  & CAHA \\
           &             &       &     &     & 1745  & CAHA\\
TY Oph     & C5,5        & Lb    & \n  &  \n & 1015  & CAHA \\
U Cam      & C3,9        & SRb   & 400 & -4.8&  791  & ORM \\
U Hya      & C6,3        & SRb   & 450 & -4.0&  947  & ORM \\
           &             &       &     &     & 1919  & ORM \\
U Lyr      & C4,5        & Mira  & 457 & -5.5&  791  & CAHA \\
UU Aur     & C5,3        & SRb   & 235 & -6.0&  947  & ORM \\
UV Aql     & C6,2        & SRa   & 385 & -7.0&  947  & ORM \\
           &             &       &     &     & 1745  & CAHA \\
UX Dra     & C7,3        & SRa   & 168 & -5.6&  947  & ORM\\
V Aql      & C6,4        & SRb   & 353 & -5.0& 1745  & CAHA \\
V CrB      & C6,2        & Mira  & 357 & -7.1& 1015  & CAHA \\
           &             &       &     &     &  947  & ORM \\
V Oph      & C5,2        & Mira  & 298 & -5.9& 1015  & CAHA \\
           &             &       &     &     &  947  & ORM\\
VX Gem     & C7,2        & Mira  & 379 & -3.6&  745  & CAHA \\
           &             &       &     &     & 1919  & ORM \\
VY Uma     & C6,3        & Lb    & \n  & -4.3& 1919  & ORM \\
V429 Cyg   & C5,4        & SRa   & 164 & \n  &  745  & CAHA \\
           &             &       &     &     & 1745  & CAHA \\
V460 Cyg   & C6,3        & SRb   & 180 & -5.8& 1015  & CAHA\\
           &             &       &     &     & 1745  & CAHA \\
V781 Sgr   & N           & Lb     & \n  &  \n & 1015  & CAHA \\
W CMa      & C6,3        & Lb  & \n  & -7.3& 1919  & ORM\\
W Ori      & C5,4        & SRb   & 186 & -5.6&  791  & CAHA\\
WZ Cas     & C9,2J       & SRa   & 186 & -6.2&  745  & CAHA\\
           &             &       &     &     &  791  & ORM\\
           &             &       &     &     & 1919  & ORM\\
Z Psc      & C7,2        & SRb   & 144 & -4.1&  745  & ORM\\
\tablenotetext{1}{Data taken from the SIMBAD database.}
\tablenotetext{2}{Bolometric magnitudes are estimated from mean K
magnitudes and HIPPARCOS parallaxes (see text).}
\tablenotetext{3}{CAHA: Centro Astron\'omico Hispano-Alem\'an de
Calar Alto; ORM: Observatorio del Roque de los Muchachos.}
\enddata
\end{deluxetable}

\clearpage

\begin{deluxetable}{lccccccc}
\tablewidth{410pt} \tablenum{2} \tablecaption{DATA FOR PROGRAM
STARS} \tablehead{\colhead{Star}           & \colhead{$T_{\rm
eff}$ }   & \colhead{Ref.\tablenotemark{a}} &
\colhead{C/O}            & \colhead{$^{12}$C/$^{13}$C}&
\colhead{Ref.\tablenotemark{b}} &
\colhead{K-[12]\tablenotemark{c}} &
\colhead{K-[25]\tablenotemark{c}}} 
\startdata
AQ And & 2970       & 2  &1.02 & 30 & 1 & -1.80 & -3.16\\
AW Cyg & 2760       & 2  &1.03 & 21 & 2 & -1.59 & -2.92 \\
EL Aur & 3000       & 6  &1.07 & 50 & 2 & -1.64 & -2.83\\
HK Lyr & 2866       & 2  &1.02 & 10 & 2 & -1.92 & -3.18\\
IRC $-$10397& 2600  & 6  &1.01 & 20 & 2 &  \nodata   &  \nodata  \\
IY Hya & 2500       & 6  &1.02 & 15 & 2 &  5.61 &  \nodata  \\
LQ Cyg & 2620       & 2  &1.10 & 40 & 2 & -1.98 & -3.19\\
RV Cyg & 2600       & 1  &1.20 & 74 & 3 & -2.87 & -1.59 \\
RX Sct & 3250       & 2  &1.04 & 60 & 2 &  \nodata   &   \nodata \\
S Sct  & 2895       & 1  &1.07 & 45 & 3 & -1.92 & -3.32\\
SS Vir & 2560       & 7  &1.05 & 10 & 2 & -2.60 & -1.34 \\
SY Per & 3070       & 3  &1.02 & 43 & 2 & -1.69 & -2.86\\
SZ Sgr & 2480       & 2  &1.03 & 8  & 2 & -1.74 & -2.43\\
TT Cyg & 2825       & 7  &1.04 & 30 & 2 & -3.50 & -2.05\\
TY Oph & 2780       & 2  &1.05 & 45 & 2 & -1.89 & -2.95\\
U Cam  & 2670       & 2  &1.02 & 40 & 2 & -1.44 & -2.57 \\
U Hya  & 2825       & 1  &1.05 & 35 & 2 & -3.00 & -1.86 \\
UU Aur & 2825       & 1  &1.06 & 50 & 3 & -1.80 & -3.06\\
UV Aql & 2750       & 2  &1.005& 19 & 2 & -1.74 & -3.06\\
UX Dra & 2900       & 1  &1.05 & 26 & 2 & -2.30 & -3.58\\
V Aql  & 2610       & 1  &1.16 & 90 & 2 & -1.74 & -3.17\\
V CrB  & 2250       & 4  &1.10 & 10 & 4 &  0.02 & -1.24\\
V Oph  & 2880       & 2  &1.05 & 11 & 1 & -1.64 & -2.96\\
VX Gem & 2500       & 5  &1.01 & 60 & 2 & -1.46 & -2.93\\
VY Uma & 2855       & 1  &1.08 & 44 & 3 & -3.56 & -2.13\\
V460 Cyg&2845       & 1  &1.06 & 61 & 3 & -2.05 & -3.43\\
V781 Sgr&3160       & 2  &1.02 & 35 & 3 & -3.22 & -1.86\\
W CMa  & 2880       & 1  &1.09 & 53 & 3 & -2.92 & -1.98\\
W Ori  & 2680       & 1  &1.20 & 79 & 3 & -3.04 & -1.69\\
WZ Cas & 3140       & 3  &1.005& 4  & 3 & -3.66 & -2.39\\
Z Psc  & 2870       & 1  &1.01 & 55 & 3 & -3.55 & -2.39\\
\tablenotetext{a}{References for $T_{\rm {eff}}$: (1) Lambert et
al. (1986); (2) Ohnaka \& Tsuji (1996); (3) Dyck, van Belle, \&
Benson (1996); (4) Kipper (1998); (5) Groenewegen et al. (1998);
(6) For these stars we have estimated the effective temperature
by comparing their spectra  with the temperature sequence
spectral atlas for C-stars created by Barnbaum et al. (1996); (7)
Bergeat, Knapik, \& Rutily (2001)} 
\tablenotetext{b}{Source for the
$^{12}$C/$^{13}$C ratio: (1) Ohnaka \& Tsuji (1996); (2) Derived
in this work or Paper II; (3) Lambert et al. (1986); (4) Kipper
(1998).} 
\tablenotetext{c}{The 12 and 25 $\mu$m fluxes are taken
from the second edition of the IRAS PSC anf the K magnitudes from
the Two-Micron Sky Survey indicated in Claussen et al. (1987).}
\enddata
\end{deluxetable}

\clearpage

\begin{deluxetable}{lcccc}
\footnotesize 
\tablenum{3} 
\tablecolumns{5} 
\tablewidth{0pt}
\tablecaption{SPECTROSCOPIC DATA} 
\tablehead{ \colhead {Element}
& \colhead{$\lambda(\rm{\AA})$} & \colhead{$\chi\rm{(eV)}$}  &
\colhead{log $gf$} & \colhead{Reference}\tablenotemark{a}}
\startdata
Ba II & 4934.10& 0.00&$-$0.150\tablenotemark{b}&1\\
La II & 4804.04& 0.23&$-$1.530                 &5 \\
La II & 4809.02& 0.23&$-$1.267                 &2\\
La II & 4748.73& 0.93&$-$0.600                 &2 \\
Ce II & 4773.96& 0.92& 0.010                   &3 \\
Ce I  & 4820.00& 0.27& $-$0.559                &4 \\
Ce I  & 4820.61& 0.17& $-$0.866                &4\\
Ce I  & 4822.54& 0.00& $-$0.420                &4\\
Nd II & 4811.35& 0.06& $-$0.570                &s\\
Nd II & 4817.18& 0.74& $-$0.760                &5\\
Nd II & 4825.48& 0.18& $-$0.880                &5\\
Nd II & 4832.27& 0.56& $-$0.920                &5\\
Sm II & 4777.85& 0.04& $-$1.340                &5\\
Sm II & 4815.80& 0.18& $-$0.770                &5\\
\tablenotetext{a}{(1) McWilliam (1998);(2) Bard, Barisciano, \&
Cowley (1996);(3) Th\'evenin (1989);(4) Meggers, Corliss, \&
Scribner (1975);(5) VALD data base (Piskunov et al. 1995);(s)
solar $gf$ derived.} 
\tablenotetext{b}{Hyperfine structure
included according to McWilliam (1998).}
\enddata
\end{deluxetable}

\clearpage

\begin{deluxetable}{cccc}
\tablenum{6} 
\scriptsize 
\tablecolumns{4} 
\tablewidth{0pt}
\tablecaption{PREDICTED C/O RATIO AND $f$ VALUE IN EXTRINSIC AGB
STARS}
\tablehead{\colhead{} & \multicolumn{3}{c}{1.5 $M_{\odot}$} \\
\colhead{[Fe/H]} &\colhead{(C/O)$_{\rm AGB}^{\rm lastTDU}$}
&\colhead{}& \colhead{$f$} } \startdata
$ +0.00 $ &     1.45 &  &  1.60  \\
$ -0.12 $ &     1.65 &  &  1.15  \\
$ -0.30 $ &     2.00 &  &  0.79  \\
$ -0.40 $ &     4.28 &  &  0.25  \\
$ -0.52 $ &     5.02 &  &  0.21  \\
$ -0.60 $ &     5.55 &  &  0.18  \\
$ -0.70 $ &     6.28 &  &  0.16  \\
$ -0.82 $ &     7.36 &  &  0.14  \\
$ -1.00 $ &     9.18 &  &  0.11  \\
$ -1.30 $ &    13.29 &  &  0.07  \\
$ -1.60 $ &    33.08 &  &  0.03  \\
\enddata
\end{deluxetable}

\clearpage

\figcaption[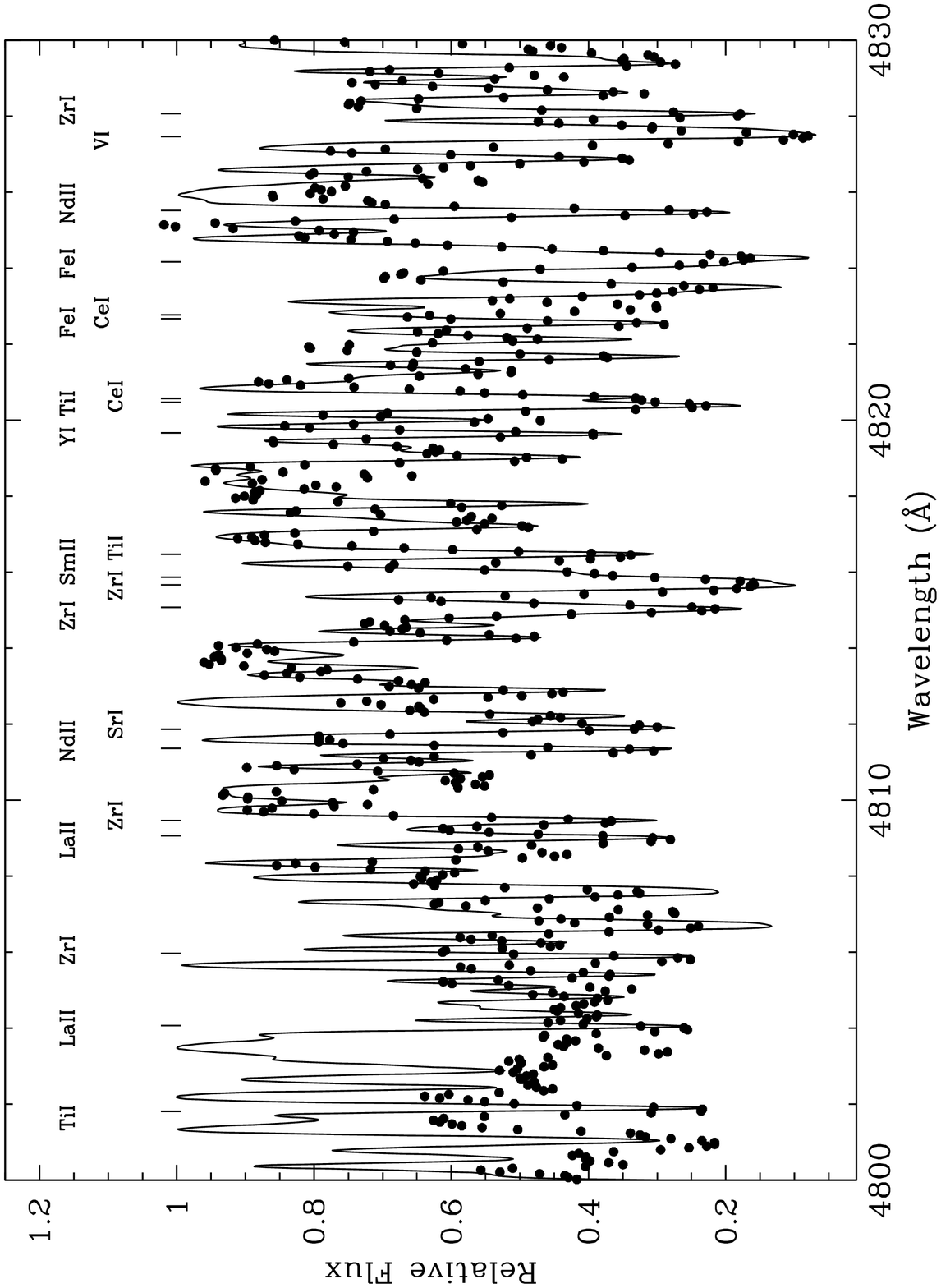]{Best  synthetic   fit  (solid  line) to
the observed  (filled circles) spectrum  of the  star Z Psc in
the range $\lambda 4800-4830$   {\AA}.  The main atomic
absorption lines  are indicated. Note that  there are still many
spectral  features not well reproduced by our synthetic
spectrum.\label{fig1}}

\figcaption[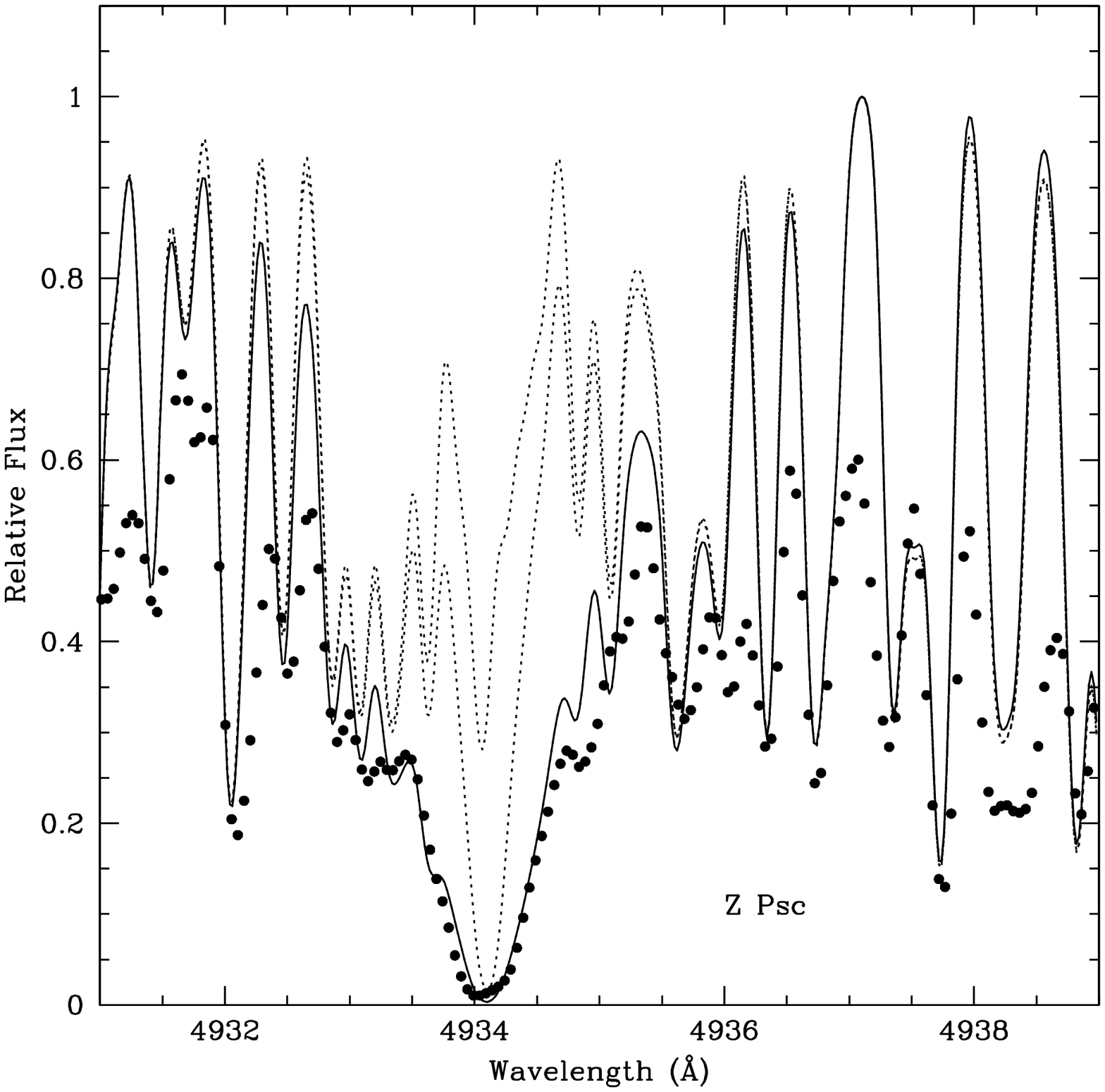]{As Figure 1 for the  star in the region
of the Ba II line at $\lambda 4934$ {\AA} for different Ba
abundances: no Ba (upper dashed line), 12 + log (Ba/H)=2.13 and 3.1 
(best fit, solid line).\label{fig2}}

\figcaption[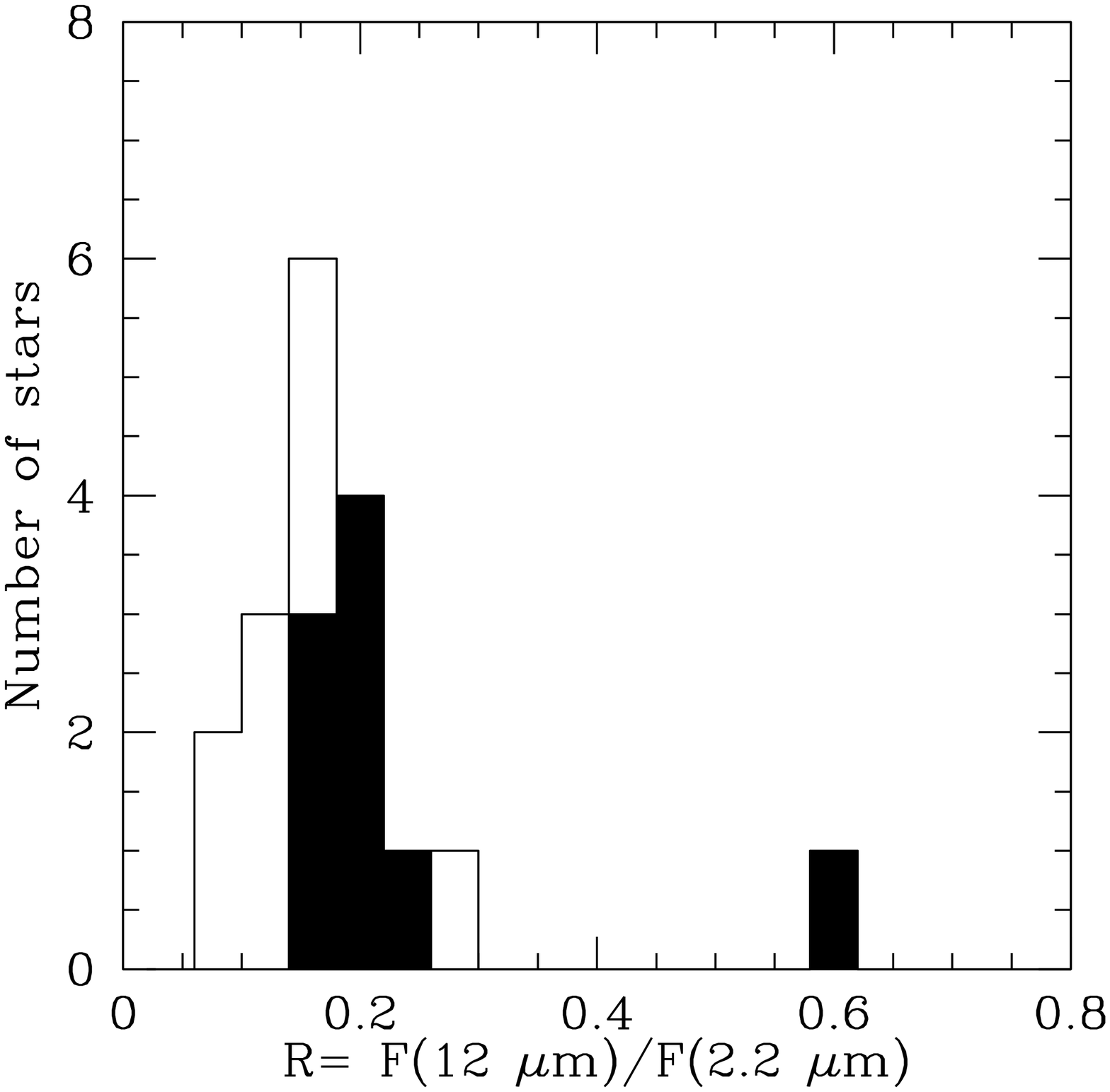]{Distribution of $R=F(12\mu m)/F(2.2\mu
m)$ for Tc-yes (open  histogram) and Tc-no (shaded  histogram) N
stars. The Tc-no stars with $R>0.6$ are collected in the last
bin.\label{fig3}}

\figcaption[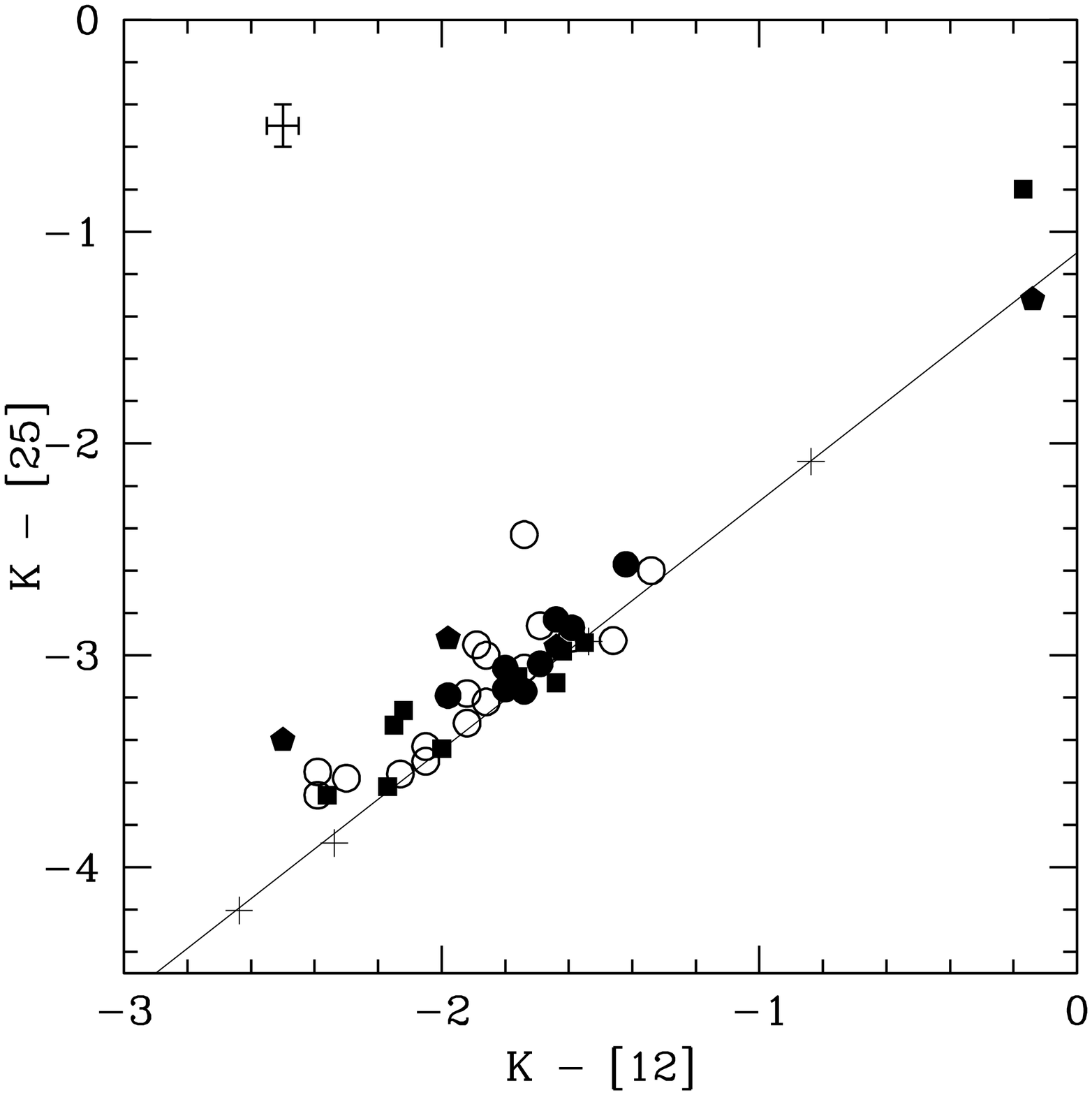]{The  (K-[12],   K-[25])  color-color
diagram, where  open circles  stand for  Tc-yes  and filled
circles for  Tc-no N stars.  Filled  diamonds  are  stars  in
our  sample  with  no information   about    Tc   (whether from
this   work    or   the literature). Filled squares are the
carbon stars of J-type studied by Abia \& Isern (2000). J-stars
do  not show Tc.  The solid  line represents black body  colors,
with  crosses corresponding to  temperatures 4000, 3000, 2000 and
1500 K.  The error bars represent a typical uncertainty of 5  and
10$\%$ on  the 12 and  25 $\mu$m fluxes,  respectively. Note that
there is  not any segregation between different  groups of carbon
stars in this diagram.\label{fig4}}

\figcaption[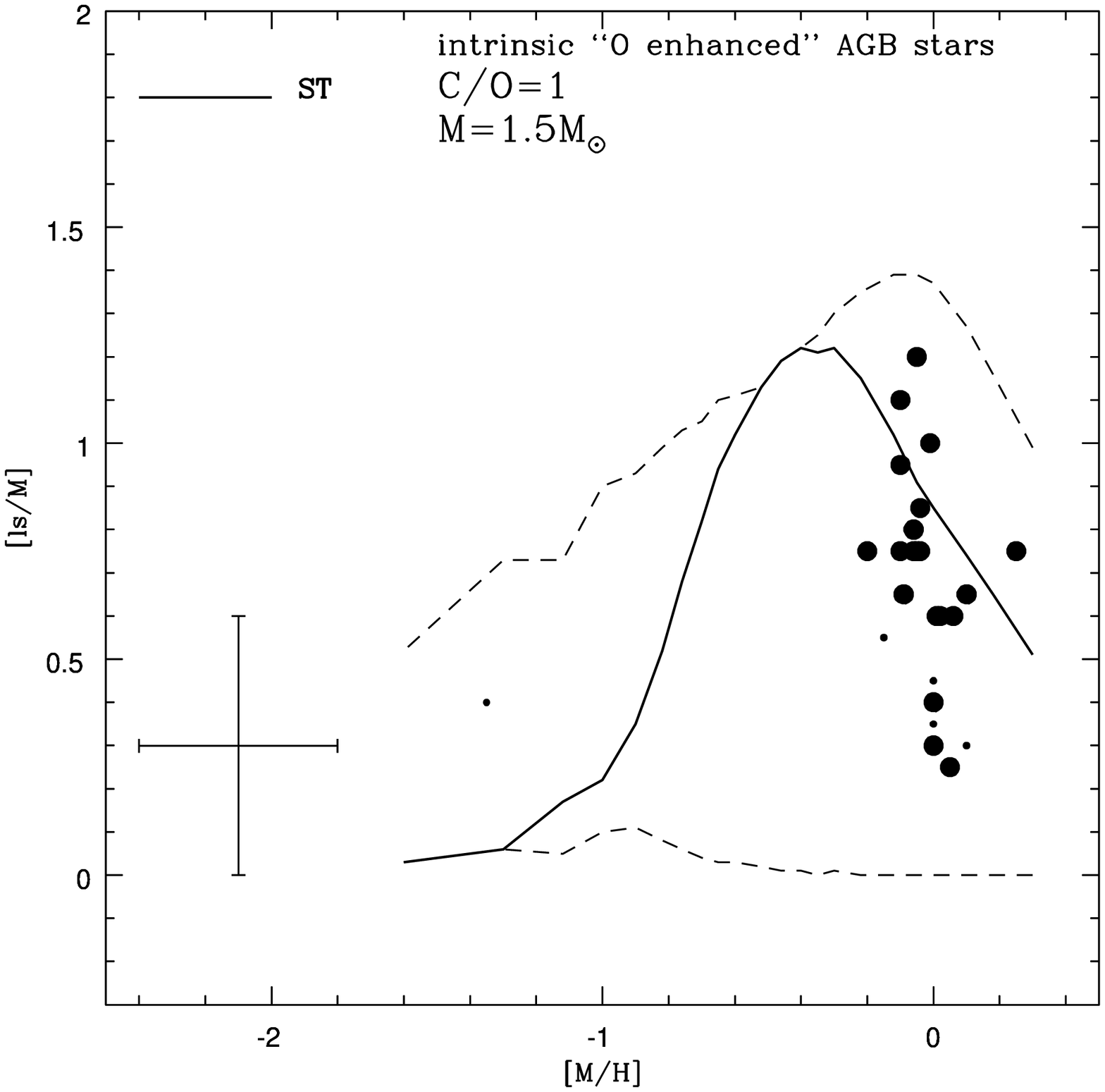]{Comparison  of   the  observed  mean
low-mass $s$-element   (Y,Zr)  [ls/M]   enhancement   against
metallicity   with theoretical predictions for a 1.5 \msb TP-AGB
star. The theoretical  predictions shown  are for O-enhanced
stellar models at  C/O$=1$. The upper and lower curves (dashed
lines) limit the region allowed by the models according to the
different $^{13}$C-pocket choices (see text). Note that  several
stars coincide in  the  same data point. The  {\it   difficult}
stars are   plotted  with   smaller symbols (see
text).\label{fig5}}

\figcaption[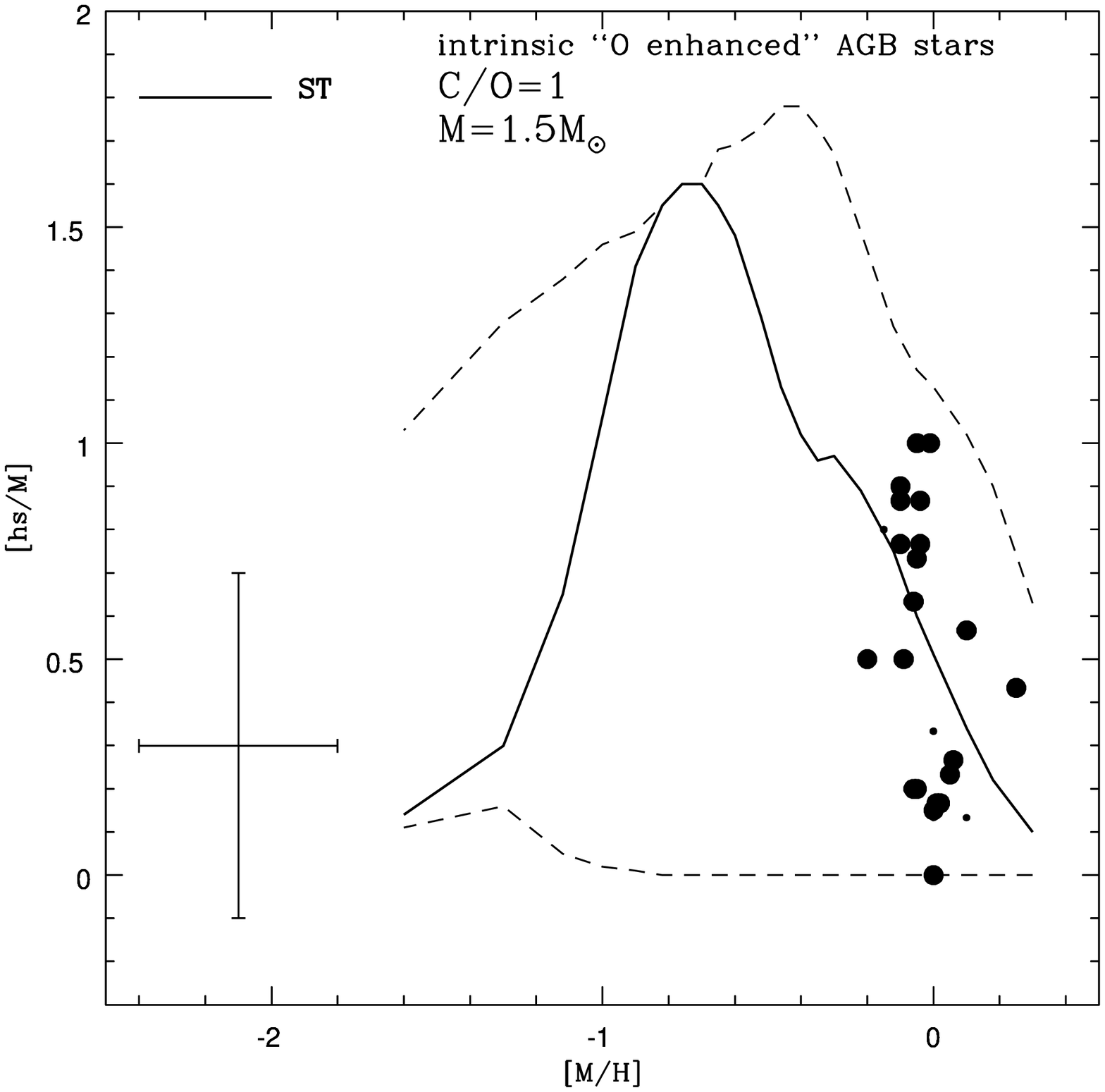]{Same  as  Figure   5  for  the  observed
mean high-mass $s$-element (Ba,La,Nd) [hs/M]  enhancement as
compared with theoretical predictions for a 1.5 \msb TP-AGB
star.\label{fig6}}

\figcaption[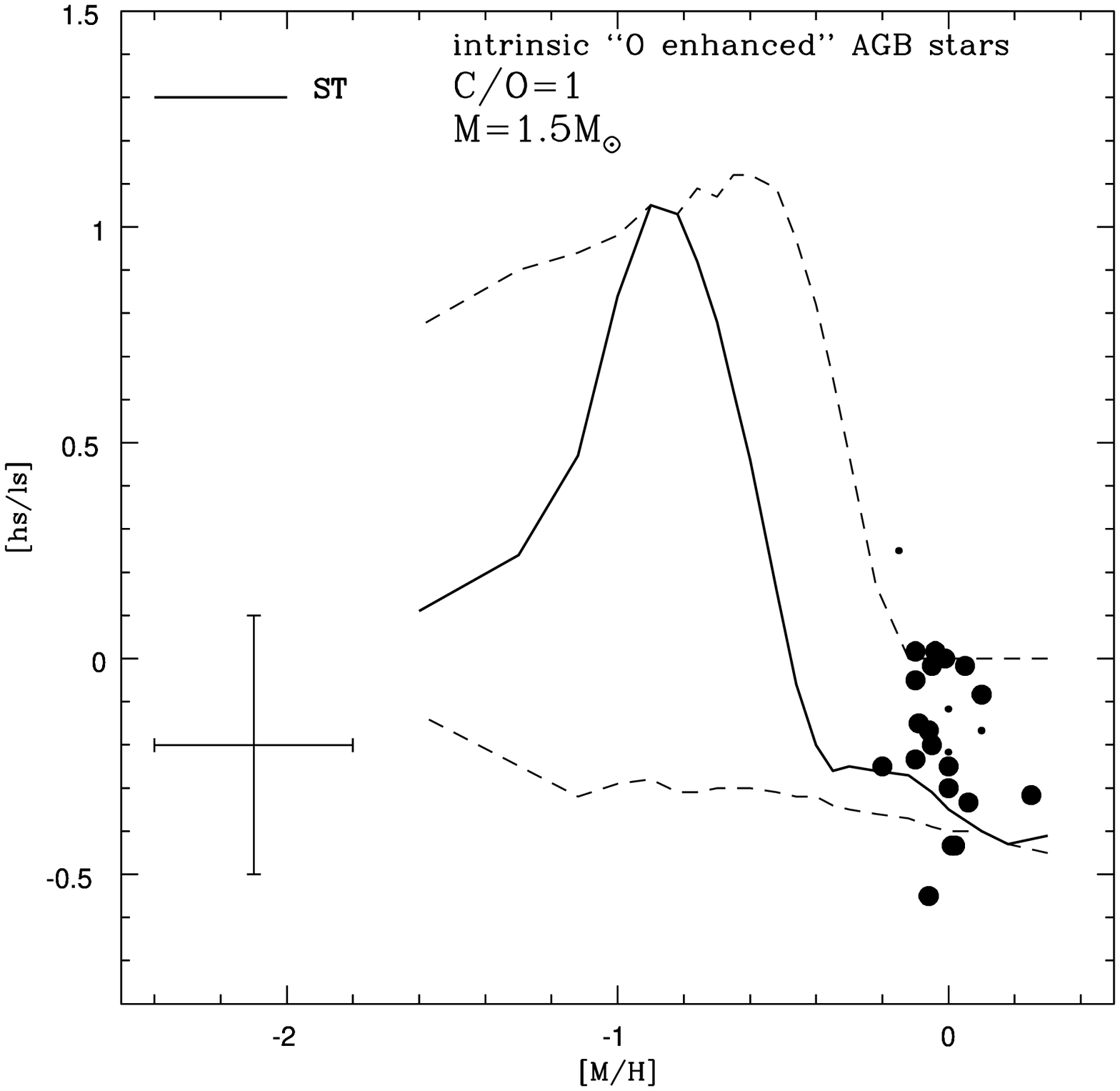]{Same  as  Figure 5  for  the observed
mean [hs/ls]  ratio (signature  of the  neutron exposure)  as
compared with theoretical predictions for a 1.5 \msb 
TP-AGB star.\label{fig7}}

\figcaption[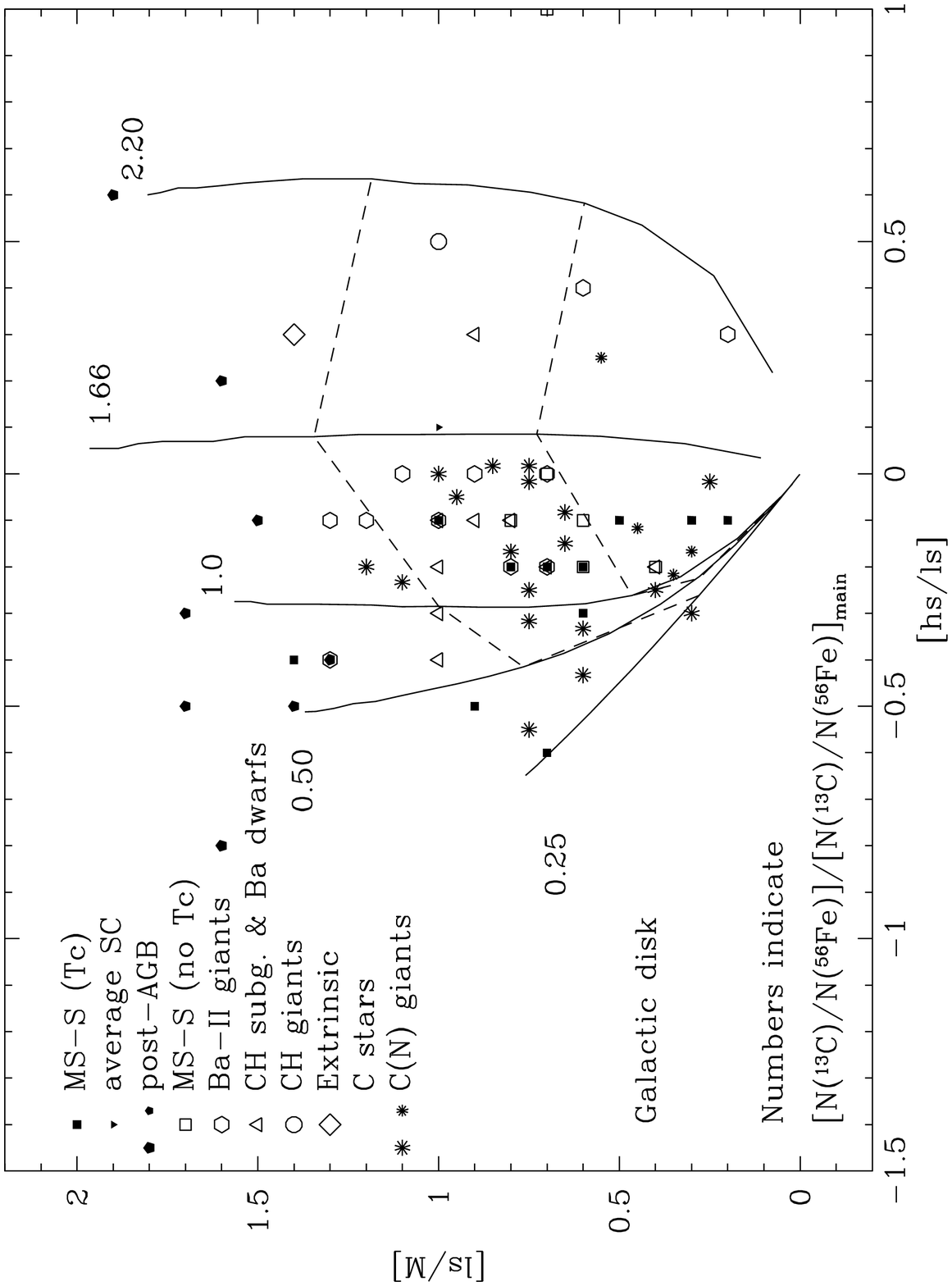]{Observed trend of the ls element
abundances [ls/M] vs. [hs/ls] for Galactic disk intrinsic and
extrinsinc AGB stars. The solid curves refer to envelope models
with different $s$-process efficiency, as monitored by the
$N(^{13}\rm{C})/N(^{56}\rm{Fe})$ ratio (here normalized to the
case that fits the main componet in the solar system). The dashed
lines connect the points corresponding to the fourth and eighth
dredge-up episode to make clear how the stars distribute along
the TP-AGB evolutionary sequence. Again, our {\it difficult} N
stars have been plotted with smaller symbols.\label{fig8}}

\figcaption[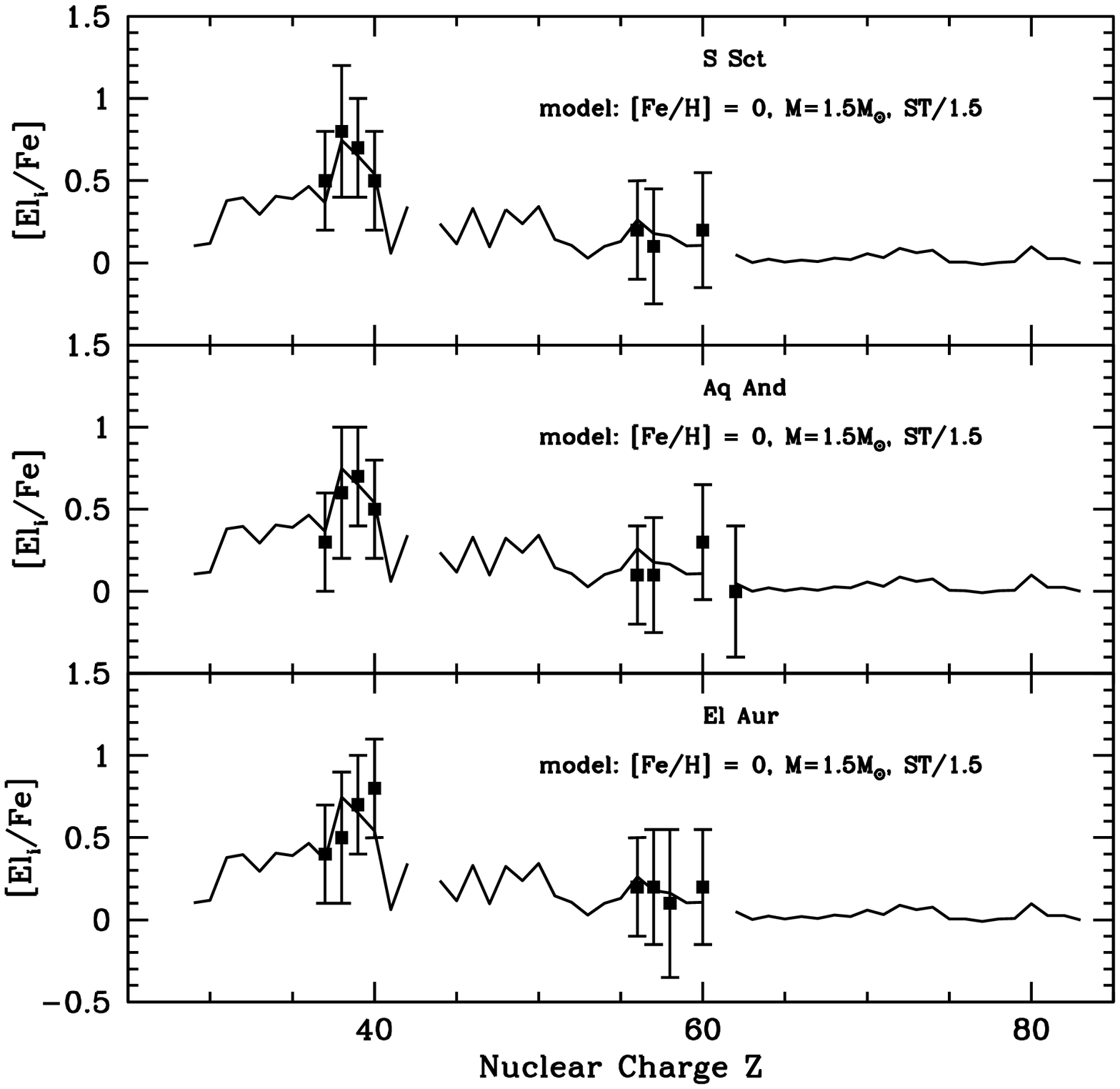]{A detailed  reproduction  of  the
observed abundances in the carbon stars S Sct, AQ And, EL Aur
(left panel), UV Aql, Z Psc, and UU Aur (right panel). See text
for details. \label{fig9}}

\plotone{f1.eps}

\plotone{f2.eps}

\plotone{f3.eps}

\plotone{f4.eps}

\plotone{f5.eps}

\plotone{f6.eps}

\plotone{f7.eps}

\plotone{f8.eps}

\plotone{f9a.eps}

\plotone{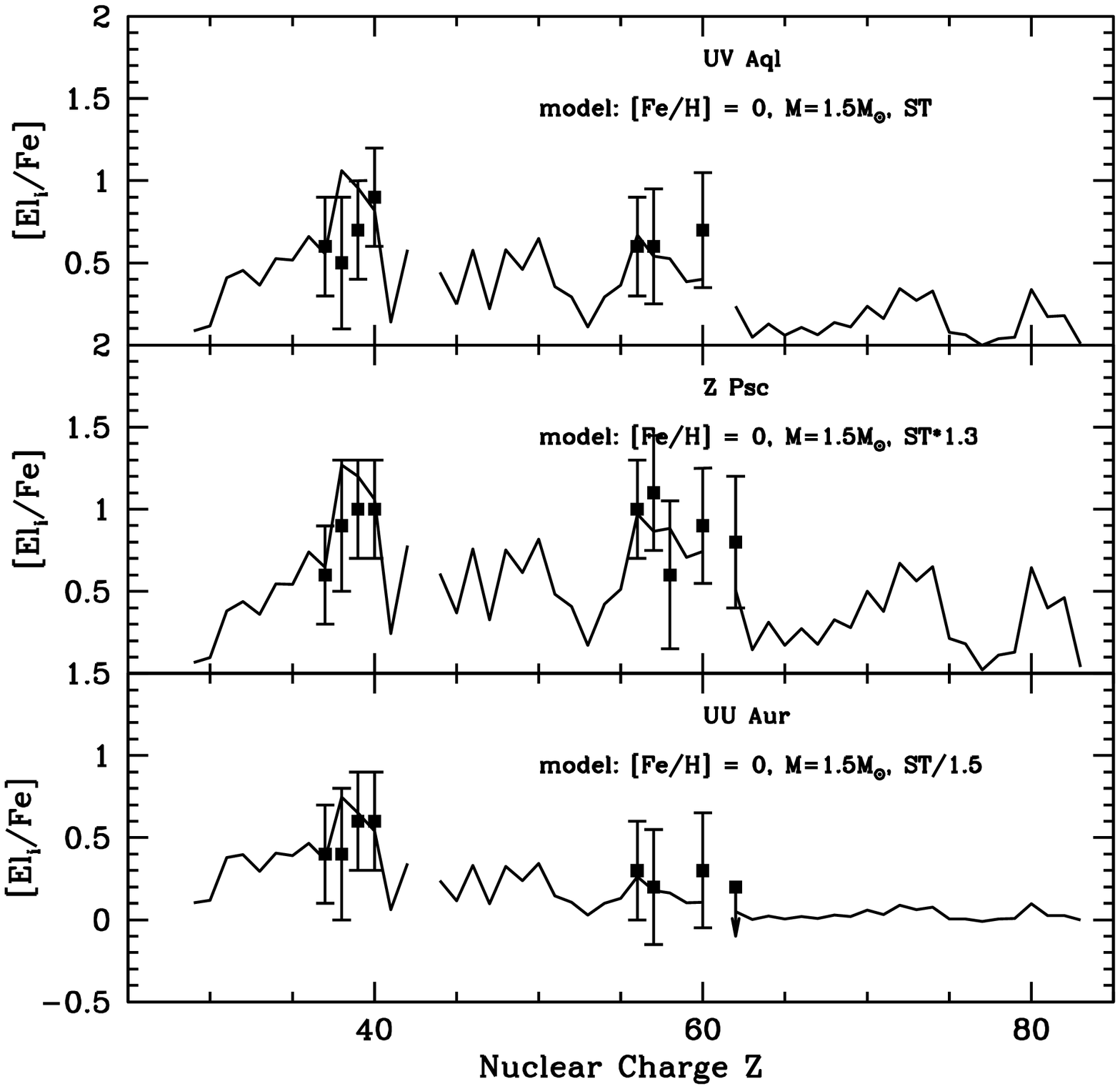}


\begin{thebibliography}{}

\bibitem[Abia \& Wallerstein  (1998)]{abi98} Abia, C., \& Wallerstein,
G. 1998, \mnras, 293, 89 (Paper I)
\bibitem[Abia  \& Isern (2000)]{abi00}  Abia, C.,  \& Isern,  J. 2000,
\apj, 449, 438
\bibitem[Abia et al. (2001)]{abi01}  Abia, C., Busso, M., Gallino, R.,
Dom\'\i nguez, I.,  Straniero, O., \& Isern, J.  2001, \apj, 559,
1117 (Paper II)
\bibitem[Alkniss  et  al. (1998)]{alk98}  Alkniss,  A., Balnauss,  A.,
Dzervitis, V., \& Eglitis, I. 1998, \aap, 338, 209
\bibitem[Anders  \& Grevesse (1989)]{and89}  Anders, E.,  \& Grevesse,
N. 1989, Geochim. Cosmochim. Acta, 53, 197
\bibitem[Arenou, Grenon, \&  G\'omez (1992)]{are92} Arenou, F., Grenon,
M., \& G\'omez, A. 1992, \aap, 258, 108
\bibitem[Bard,  Barisciano,   \&  Cowley  (1996)]{bard96}   Bard,  D.J.,
Barisciano, L.P., \& Cowley, C.R. 1996, \mnras, 278, 997
\bibitem[Barnbaum, Stone, \& Keenan (1996)]{barn96}Barnbaum, C., Stone,
R.P.S., \& Keenan, P.C. 1996, \apjs, 105, 419
\bibitem[Bauschlicher, Langhoff, \& Taylor (1988)]{bau88} Bauschlicher,
C.W., Langhoff, S.R., \& Taylor, P.R. 1988, \apj, 332, 531
\bibitem[Beer  \&   Macklin  (1989)]{bee89}  Beer,   H.,  \&  Macklin,
R. L. 1989, \apj, 339, 962
\bibitem[Bergeat, Knapik, \&  Rutily (2001)]{ber01} Bergeat, J., Knapik,
A., \& Rutily, B. 2001, \aap, 369, 178
\bibitem[Bergeat,  Knapik,  \&   Rutily  (2002a)]{ber02a}  Bergeat,  J.,
Knapik, A., \& Rutily, B. 2002a, \aap, 385, 94
\bibitem[Bergeat,  Knapik,  \&   Rutily  (2002b)]{ber02b}  Bergeat,  J.,
Knapik, A., \& Rutily, B. 2002b, \aap, (in press)
\bibitem[Beveridge \& Sneden (1994)]{bs94} Beveridge, R.C., \& Sneden,
C. 1994, \aj, 108, 265
\bibitem[Busso et al. (1995)]{bus95} Busso, M., Lambert, D.L., Beglio,
L., \& Gallino, R. 1995, \apj, 446, 775
\bibitem[Busso,  Gallino,  \&  Wasserburg  (1999)]{bus99}  Busso,  M.,
Gallino, R., \& Wasserburg, G.J. 1999, \araa, 37, 239
\bibitem[Busso  et  al.  (2001a)]{bus01}  Busso,  M.,  Lambert,  D.L.,
Gallino, R., Travaglio, C., \& Smith, V.V. 2001a, \apj, 557, 802
\bibitem[Busso   et   al.   (2001b)]{bus01b}   Busso,   M.,   Marengo,
M.,Travaglio, C.,  Corcione, L., \& Silvestro,  G. 2001b, Mem. Soc. 
Astron. It., 72, 309
\bibitem[Cardelli,  Clayton, \&  Mathis (1989)]{car89}  Cardelli, J.A.,
Clayton, G.C., \& Mathis, J.S. 1989, \apj, 345, 245
\bibitem[Cerny et al. (1978)]{cer78}Cerny, D., Bacis, R., Guelachvili,
G., \& Roux, F., 1978, JMS, 73, 154
\bibitem[Claussen  et al.  (1987)]{cla87}  Claussen, M.J.,  Kleinmann,
S.G., Joyce, R.R., \& Jura, M. 1987, \apjs, 65, 385
\bibitem[Corliss  \& Bozman (1962)]{cor62}  Corliss, C.H.,  \& Bozman,
W.R. 1962, NBS Monograph 53
\bibitem[de Laverny  \& Gustafsson  (1998)]{deL98} de Laverny,  P., \&
Gustafsson, B. 1998, \aap, 332, 661
\bibitem[Dominy (1985)]{dom85} Dominy, J. 1985, \pasp, 97, 1104
\bibitem[Dyck,  van Belle,  \& Benson  (1996)]{dyc96} Dyck,  H.M., van
Belle, G.T., \& Benson, J.A. 1996, \aj, 112, 96
\bibitem[Epchtein  et  al.  (1999)]{epch99}  Epchtein, N.,  Deul,  E.,
Derriere,  S., Borsenberger,  J.,  Egret, D.,  Simon,  G.,
Alard,  C., Bal\'azs, L. et al. 1999, \aap, 349, 236
\bibitem[Eriksson et al.  (1984)]{eri84} Eriksson, K., Gustafsson, B.,
J\o rgensen, U., \& Nordlund, A. 1984, \aap, 132, 37
\bibitem[Gallino  et al.  (1998)]{gal98} Gallino,  R.,  Arlandini, C.,
Busso, M., Lugaro,  M., Travaglio, C., Straniero, O.,  Chieffi,
A., \& Limongi, M. 1998, \apj, 497, 388
\bibitem[Gezari  et  al.  (1999)]{gez99}  Gezari, D.Y.,  Pitts,  P.S.,
\& Schmitz, M.  1999,  Catalog   of  Infrared  Observations, 5th
edition, electronic release at the SIMBAD database, CDS,
Strasbourg.
\bibitem[Groenewegen   et  al.  (1989)]{gro89}   Groenewegen,  M.A.T.,
Whitelock, P.A., Smith, C.H., \& Kerschbaum, F. 1998, \mnras,
293, 18
\bibitem[Gustafsson \& J\o  rgenssen (1994)]{gus94} Gustafsson, B., \&
J\o rgenssen, U.G. 1994, \aap Rev., 6, 19
\bibitem[Holweger   \&  M\"uller   (1974)]{how74}  Holweger,   H.,  \&
M\"uller, E.A. 1974, Sol. Phys., 39, 19
\bibitem[Iben  \&  Renzini (1983)]{ibe83}  Iben,  I.  Jr, \&  Renzini,
A. 1983, \araa, 21, 271
\bibitem[Israelian et al. (2001)]{isr01} Israelian, G., Rebolo, R.,
Garc\'\i a L\'opez, R., Bonifacio, P., Molaro, P., Basri, G., \&
Shchukina, N. 2001, \apj, 551, 833
\bibitem[Ito  et al.  (1988)]{ito88}Ito,  H., Ozaki,  Y., Suzuki,  K.,
Kondow, T., \& Kuchitsu, K., 1988, JMS, 127, 283
\bibitem[J\o rgensen et  al. (1996)]{jor96}J\o rgensen, U.G., Larsson,
M., Iwamae, A., \& Yu, B., 1996, \aap, 315, 204
\bibitem[Jorissen \&  Mayor  (1988)]{jm88}  Jorissen, A.,  \&  Mayor,
M. 1988, A\&A, 198, 187
\bibitem[Jorissen et al. (1993)]{jor93}  Jorissen, A.,  Frayer, D.T.,
Johnson, H.R., Mayor, M., \& Smith, V.V. 1993, \aap, 271, 463
\bibitem[Jura (1986)]{jur86}Jura, M. 1986, Irish Astr. J., 17, 332
\bibitem[Jura  \&  Kleinmann (1989)]{jur89}  Jura,  M., \&  Kleinmann,
S.G. 1989, \apj, 341, 359
\bibitem[Knapp  &  Morris   (1985)]{kna85}  Knapp,  G.R.,  \&  Morris,
M. 1985, \apj, 292, 640
\bibitem[Knapp,  Pourbaix, \&  Jorissen (2001)]{kna001} Knapp,  G.R.,
Pourbaix, D., \& Jorissen, A.  2001, \aap, 371, 222
\bibitem[Kilston (1975)]{kil75}Kilston, S. 1975, \pasp, 87, 109
\bibitem[Kipper (1998)]{kip98} Kipper, T. 1998, Baltic Astronomy A, 7,
435
\bibitem[Kotlar, Field, \& Steinfeld (1980)]{kot80}Kotlar, A.J., Field,
R.W., \& Steinfeld, J.I., 1980, JMS 80,86
\bibitem[Kupka et al. (1998)]{kup98}Kupka F., Piskunov N., Ryabchikova
T.A., Stempels H.C., \& Weiss W.W., 1998, \aaps, 138, 119
\bibitem[Lambert et al.  (1986)]{lam86} Lambert, D.L., Gustafsson, B.,
Eriksson, K., \& Hinkle, K. H. 1986, \apjs, 62, 373
\bibitem[Larsson,  Siegbahn, \&  \AA gren  (1983)]{lar83}Larsson,  M.,
Siegbahn, P.E.M., \& \AA gren, H., 1983, \apj, 272, 369
\bibitem[Little-Marenin \& Little (1988)]{lit88} Little-Marenin, I.R.,
\& Little, S.J. 1988, \apj, 333, 305
\bibitem[Lloyd-Evans (1983)]{llo83}Lloyd-Evans,  T. 1983, \mnras, 204,
975
\bibitem[Luck &  Bond (1991)]{luc91} Luck,  R.E., \& Bond,  H.E. 1991,
\apjs, 77, 515
\bibitem[Luque  \&   Crosley  (1999)]{lif99}Luque,  J.,   \&  Crosley,
D.R. 1999 SRI International Report MP 99-009
\bibitem[McWilliam (1998)]{mac98}McWilliam, A. 1998, \aj, 115, 1648
\bibitem[Marengo et al. (1999)]{mar99}  Marengo, M., Busso, M., Persi,
P., Silvestro, G., \& Lagage, P.O. 1999, \aap, 348, 501
\bibitem[Marengo, Ivezic, \& Knapp (2001)]{mar01} Marengo, M.,
Ivezic, Z., \& Knapp, G.R. 2001, \mnras, 324, 111
\bibitem[Meggers,  Corliss, \&  Scribner (1975)]{mer75}  Meggers, W.F.,
Corliss, C.H., \& Scribner, B.F. 1975, NBS Monograph 145
\bibitem[Merill (1952)]{mer52} Merill, P.W. 1952, \apj, 116, 21
\bibitem[Neugebauer  \&  Leighton  (1969)]{neu69} Neugebauer,  G.,  \&
Leighton, R.B. 1969, Two-Micron Sky Survey, NASA SP-3047(TMSS)
\bibitem[Nollett,  Busso, \& Wasserburg  (2002)]{nol02} Nollett,  K. M.,
Busso, M., \& Wasserburg, G. 2002, \apj, in press
\bibitem[Ohnaka \& Tsuji (1996)]{ohn96} Ohnaka, K., \& Tsuji, T. 1996,
\aap, 310, 933
\bibitem[Olofsson  et  al.   (1993)]{olo93}  Olofsson,  H.,  Eriksson,
K. Gustafsson, B., \& Carlstrom, U. 1993, \aaps, 87, 267
\bibitem[Paczy\'nsky   (1970)]{pac70}  Paczy\'nsky,   B.   1970,  Acta
Astron., 20, 47
\bibitem[Piskunov  et al.  (1995)]{pis95} Piskunov,  N.E.,  Kupka, F.,
Ryabchikove, T.A., Weiss, W.W., \& Jeffry, C.S. 1995, \aaps, 285,
541
\bibitem[Plez (1998)]{ple98} Plez, B. 1998, \aap, 337, 495
\bibitem[Pourbaix,  Knapp, \&  Jorissen (2002)]{pou02} Pourbaix,  D.,
Knapp, G.R., \& Jorissen, A. 2002, \aap, (submitted)
\bibitem[Prasad \& Bernath  (1992)]{pra92} Prasad, C.V.V., \& Bernath,
P.F., 1992, JMS 156, 327
\bibitem[Prasad et al.  (1992)]{pra92b} Prasad, C.V.V., Bernath, P.F.,
Frum, C., \& Engleman, R.Jr., 1992, JMS 151, 459
\bibitem[Reddy, Bakker, \&  Hrivnak (1999)]{red99} Reddy, B.E., Bakker,
E.J, \& Hrivnak, B.J. 1999, \apj, 524, 831
\bibitem[Reddy et al. (2002)]{red02} Reddy, B.E., Lambert, D.L., Gonzalez, G.,
\& Yong, D. 2002, \apj, 564, 482
\bibitem[Rehfuss,  Suh, \&  Miller  (1992)]{reh92}Rehfuss, B.D.,  Suh,
M.H., \& Miller, T.A., 1992, JMS 151, 43
\bibitem[Reimers  (1975)]{rei75}  Reimers,  D.  1975, In  Problems  in
Stellar  Atmospheres  and  Envelopes,  ed.   B.  Bascheck,  H.
Kegel, G. Traving, p. 229. Berlin: Springer-Verlag
\bibitem[Schwarzschild  \& H\"arm  (1965)]{sch65} Schwarzschild,  M., \&
H\"arm, R. 1965, \apj, 145, 496
\bibitem[Skrutskie  et al.  (1997)]{skr97} Skrutskie,  M.F., Scheider,
S.E.,  Stiening,  R.,  Strom,  S.E.,  Weinberg,  M.D., Beichman,
C., Chester, T.,  \& Cutri, R., et al.  1997, in Proceedings  of
the Workshop 'The Impact of Large Scale Near-Infrared Surveys',
p. 25
\bibitem[Smith  \&  Lambert (1985)]{smi85}  Smith,  V.V., \&  Lambert,
D.L. 1985, \apj, 294, 326
\bibitem[Smith  \&  Lambert (1986)]{smi86}  Smith,  V.V., \&  Lambert,
D.L. 1986, \apj, 311, 843
\bibitem[Smith  \&  Lambert (1990)]{smi90}  Smith,  V.V., \&  Lambert,
D.L. 1990, \apjs, 72, 387
\bibitem[Straniero et  al. (1995)]{str95} Straniero,  O., Gallino, R.,
Busso, M., Chieffi, A., Raiteri,  C., Salaris, M., \& Limongi, M.
1995, \apj, 440, L85
\bibitem[Straniero et  al. (1997)]{str97} Straniero,  O., Chieffi, A.,
Limongi, M. Busso, M., \& Gallino, R.  1997, \apj, 478, 332
\bibitem[Straniero et al. (2000)]{str00} Straniero, O., Limongi, M., 
Chieffi, A., Dom\'\i nguez, I., Busso, M., \& Gallino, R. 2000, Mem. Soc.
Astron. It., 71, 719
\bibitem[Th\'evenin (1989)]{the89} Th\'evenin, F. 1989, \aaps, 77, 137
\bibitem[Th\'evenin (1990)]{the90} Th\'evenin, F. 1990, \aaps, 82, 179
\bibitem[Tomkin  \&  Lambert (1983)]{tom83}  Tomkin,  J., \&  Lambert,
D.L. 1983, \apj, 273, 722
\bibitem[Utsumi (1970)]{uts70} Utsumi, K. 1970, \pasj, 22, 93
\bibitem[Utsumi  (1985)]{uts85} Utsumi,  K. 1985,  in Cool  Stars with
Excesses of Heavy Elements, eds. M. Jascheck \& P.C.Keenan
(Dordrecht: Reidel), 243
\bibitem[Vanture (1992a)]{van92a} Vanture, A.D. 1992a, \aj, 103, 2035
\bibitem[Vanture (1992b)]{van92b} Vanture, A.D. 1992b, \aj, 104, 1997
\bibitem[Vanture (1992c)]{van92c} Vanture, A.D. 1992c, \aj, 104, 1986
\bibitem[van Leeuwen \& Evans  (1998)]{van98}van Leeuwen, F., \& Evans,
D.W. 1988, \aap, 130, 157
\bibitem[Van  Winckel \&  Reyniers (2000)]{van00}Van  Winckel,  H., \&
Reyniers, C. 2000, \aap, 354, 135
\bibitem[Wallerstein  et   al.  (1997)]{wal97}  Wallerstein,   G.,  et
al. 1997, Rev. Mod. Phys., 69, 4
\bibitem[Wallerstein  \&  Knapp  (1998)]{wal98}  Wallerstein,  G.,  \&
Knapp, G.R. 1998, \araa, 36, 369
\bibitem[Wasserburg et al. (1995)]{was95} Wasserburg, G.J., Boothroyd,
A.I., \& Sackmann, I.-J. 1995, \apj, 440, L101
\bibitem[Westerlund et al. (1991)]{wes91} Westerlund, B.E., Azzopardi,
M., Breysacher, J., \& Rebeirot, E. 1991, \aap, 244, 367
\bibitem[Westerlund et al. (1995)]{wes95} Westerlund, B.E., Azzopardi,
M., Breysacher, J., \& Rebeirot, E. 1995, \aap, 303, 107
\end{thebibliography}
\end{document}